\documentclass[iop]{emulateapj}

\shorttitle{Detectability of visible-wavelength line emission from the local CGM and IGM}
\shortauthors{Lokhorst et al.}

\begin{document}
\title{On the detectability of visible-wavelength line emission from the local circumgalactic and intergalactic medium} 

   \author{Deborah Lokhorst\altaffilmark{1,2,5}, Roberto Abraham\altaffilmark{1,2}, Pieter van Dokkum\altaffilmark{3}, Nastasha Wijers\altaffilmark{4}, Joop Schaye\altaffilmark{4}    }

\altaffiltext{1}{Department of Astronomy \& Astrophysics, University of Toronto, 50 St. George Street, Toronto, ON, M5S 3H4 }
\altaffiltext{2}{Dunlap Institute of Astronomy \& Astrophysics, University of Toronto, 50 St. George Street, Toronto, ON, M5S 3H4 }
\altaffiltext{3}{Department of Astronomy, Yale University, 260 Whitney Avenue, New Haven, CT 06511 }
\altaffiltext{4}{Leiden Observatory, Leiden University, P.O. Box 9513, 2300 RA Leiden, The Netherlands}
\altaffiltext{5}{Email address: lokhorst@astro.utoronto.ca}

\begin{abstract}

We describe a new approach to studying the intergalactic and circumgalactic medium in the local Universe: direct detection through narrow-band imaging of ultra-low surface brightness visible-wavelength line emission. 
We use the hydrodynamical cosmological simulation EAGLE to investigate the expected brightness of this emission at low redshift ($z$ $\lesssim$ 0.2).  H$\alpha$ emission in extended halos (analogous to the extended Ly$\alpha$ halos/blobs detected around galaxies at high redshifts) has a surface brightness of $\gtrsim700$ photons cm$^{-2}$ sr$^{-1}$ s$^{-1}$ out to $\sim$100 kpc.  Mock observations show that the Dragonfly Telephoto Array, equipped with state-of-the-art narrow-band filters, could directly image these structures in exposure times of $\sim$10 hours.  H$\alpha$ fluorescence emission from this gas can be used to place strong constraints on the local ultra-violet background, and on gas flows around galaxies.  Detecting H$\alpha$ emission from the diffuse intergalactic medium (the ``cosmic web") is beyond current capabilities, but would be possible with a hypothetical 1000-lens Dragonfly array.

\end{abstract}


\section{Introduction}
The intergalactic medium (IGM), together with its close cousin the circumgalactic medium (CGM), are arguably the most important baryonic components of the Universe. 
The IGM is composed mainly of a diffuse plasma of primordial hydrogen and helium polluted by small quantities of metals produced by star formation. The near-invisibility of the IGM (see below) masks its absolutely fundamental importance: the IGM contains the majority of baryons in the Universe, and it is the ultimate source of fuel for the star formation occurring in galaxies \citep[see, e.g.,][for a review]{mcqu2015}. In most models, this gaseous fuel flows along the cosmic web of filamentary dark matter pervading the Universe. Galaxies form within dark matter halos at the intersections of the filaments.   As the IGM gas falls into halos, it transitions into the CGM, the physics of which are a complex interplay between the large scale dynamics of the infalling gas and feedback of reprocessed gas (and energy) back into the CGM from the galaxies themselves.  The exact definition of the CGM is still debated, but it can be roughly described as being bounded by the disk or interstellar medium of the galaxy on the inside, and the virial radius of a galaxy's dark matter halo on the outside \citep[see, e.g.,][and references therein]{tuml2017}. 

The CGM is central to building galaxies, but it is still poorly understood. The gas depletion timescale of galaxies is short (typically 1--2 Gyr), so accretion onto galaxies is necessary to sustain measured star formation rates \citep[e.g.][]{baue10,vand11}.
But beyond this, we know little about how galactic star formation is fueled by the IGM.
We do not even know basic facts such as the typical amount of gas in the CGM, or even whether this gas is at the virial temperature of the halos.  
Because of this, we also do not know the processes by which this gas is accreted onto the central galaxy. Once the gas has made it into galaxies, we do not know how much gas is blown back out again by winds. This is also important because the porosity of the gas within the CGM determines how much ultraviolet radiation from galactic star-formation leaks out into the IGM.  Therefore, the effective range over which a typical galaxy influences the ionization state of the Universe is not clear.  

Why is so much still not understood about the IGM/CGM, particularly at low redshifts?  In principle, some of the relevant physics of the CGM and IGM can be probed directly by HI imaging at 21cm or molecular gas imaging with radio telescopes, since denser pockets of the CGM are in the form of `dark' clouds of neutral hydrogen and molecular gas. Thus far this approach has met with limited success \citep[e.g.][]{oost07,heal11,moss17,varg17,ping18,emon18}. 
Single dish radio telescopes have the required sensitivity to probe cold gas in halos in the nearby Universe ($z<0.1$), but they lack the needed resolution, while radio interferometers have the required resolution but they lack the necessary dynamic range\footnote{For these reasons, the detailed investigation of the local CGM/IGM is a major goal of next-generation radio facilities, including the Square Kilometer Array.}.  Therefore, the majority of our observational constraints on the neutral components of the IGM come from studies of absorption systems. Since Ly$\alpha$ is a UV resonance line that must be cosmologically band-shifted in order to be accessible to ground-based telescopes, studies of Ly$\alpha$ absorption systems focus mainly on the characteristics of the IGM and CGM at redshifts $z>2.5$, when Ly$\alpha$ becomes band-shifted into visible wavelengths. 
Ly$\alpha$ absorption systems at lower redshifts can only be investigated using space-based UV spectroscopy, and at present the only facility available for undertaking such work is the Cosmic Origins Spectrograph (COS) on the Hubble Space Telescope \citep[see, e.g., COS-Halos and other HST-COS surveys;][]{danf16,rich16,werk13}. MgII absorption is similarly used as a tracer of neutral H column densities, observed in a redshift range that cannot be accessed by Ly$\alpha$ from the ground.  Such investigations probe the IGM in dense pockets and in pencil beams where the CGM intersects with light from background sources.

Simulations of the CGM and IGM at intermediate redshifts ($2<z<5$) have reached the point that they are now quite successful at reproducing the observed column densities of HI probed by Ly$\alpha$ absorption systems \citep[e.g.][]{alta11,rahm15}. However, discrepancies begin to occur as these simulations are advanced in time to predict the properties of the IGM in the local Universe: when one adds together the baryons contained in galaxies and those measured though Ly$\alpha$ absorption, the majority of baryons are not accounted for \citep{mcqu2015}.
Unless high redshift estimates of baryon content are incorrect, a large fraction of the low redshift baryons have been missing in observations; a significant fraction of these ``missing baryons'' are thought to exist in the warm-hot intergalactic medium (WHIM) at T $\sim10^5 - 10^7$K which is mostly invisible in Ly$\alpha$ absorption line studies \citep[see e.g.][for a review]{bert08}. 
Studies of photoionized Ly$\alpha$ and highly ionized oxygen absorbers are starting to reveal gas in the WHIM, so far constraining the total baryonic fraction of gas in the WHIM to 24 -- 40\%  \citep[e.g.][]{nica18,shul12}.
In addition, there is a current debate over the total mass of gas in the cold (T $\sim 10^4$ K) CGM of L $\sim$ L$_{\star}$ galaxies, between M$^{\mathrm{cool}}_{\mathrm{CGM}}$ $\sim3\times10^{10}$ M$_{\odot}$ \citep{keen17,stoc13} and M$^{\mathrm{cool}}_{\mathrm{CGM}}$ $\sim9\times10^{10}$ M$_{\odot}$ \citep{proc17,werk14}, which increases the uncertainty of the total cosmic baryon mass.

An exciting alternative to absorption line studies is direct imaging of emission from the IGM itself.  At $\sim$10$^5$ K, the warm-hot plasma is cooling radiatively by line emission, so the IGM is weakly luminescent at UV and visible wavelengths.
FeII and MgII emission from localized ($<$20 kpc radial distance) outflows of low redshift star-forming galaxies has been detected \citep[e.g.][]{rubi11,mart13}.
Recently, more extended Ly$\alpha$ emission from the IGM or CGM at high redshifts has begun to be investigated by spatially resolved spectrometers such as the Cosmic Web Imagers (CWIs) on Keck and Palomar \citep{mart10,matu10} and the Multi Unit Spectroscopic Explorer (MUSE) on the Very Large Telescope \citep{baco10,wiso16,wiso18,bori16}.

Ly$\alpha$ emission in low redshift galaxies is not accessible from the ground, but an appreciable fraction of the energy emitted as ultraviolet photons also emerges in visible wavelengths (such as H$\alpha$ and [OIII] 5007$\mathrm{\AA}$). 
Furthermore, these lines may be easier to interpret than Ly$\alpha$: diffuse Ly$\alpha$ emission in the outer halos of galaxies may be affected by the presence of resonantly scattered radiation suspected to originate from the central galactic HII regions \citep[see e.g.][]{stei11} and low surface brightness measurements of a non-resonant line such as H$\alpha$ can help disentangle the properties of the CGM \citep[e.g.][]{leib18}.

Measurements obtained using other hydrogen emission lines, such as H$\alpha$, may be cleaner probes of the CGM than Ly$\alpha$, but is CGM emission from these lines practically detectable?
Van de Voort \& Schaye (2013) calculated H$\alpha$ line emission from the CGM in the optically thin limit for a specified UV background and predicted that an H$\alpha$ radial profile corresponding to the Ly$\alpha$ profile observed by \citet{stei11} can be observed out to 0.2 -- 0.6 R$_{vir}$ at a surface brightness limit\footnote{This is a conservative estimate since \citet{vand13} ignored self-shielding and H$\alpha$ powered by local star formation or local fluorescence, which could significantly boost the emission.} of $10^{-20}\, {\rm erg}\, {\rm cm}^{-2}\,  {\rm s}^{-1}\, {\rm arcsec}^{-2}$.
Recent advances in low surface brightness imaging telescopes may have brought such observations into the realm of being practical.
In this paper, we investigate whether is may be possible for ground-based telescopes to observe the cooling emission from the CGM/IGM in visible wavelengths.  Our analysis is based on a subset of simulations from the Evolution \& Assembly of GALaxies \& their Environments (EAGLE) project \citep{scha15,crai15}.  We supplement the results from the simulation by calculating H$\alpha$ surface brightness estimates analytically from observations and theoretical considerations.  
We show that at low redshift, H$\alpha$ emission from diffuse structures could be targeted through an upcoming narrow-band imaging upgrade to the Dragonfly Telephoto Array (hereafter Dragonfly)\footnote{http://www.dragonflytelescope.org/}.

\begin{figure*}
\centering
\includegraphics[width=0.99\linewidth]{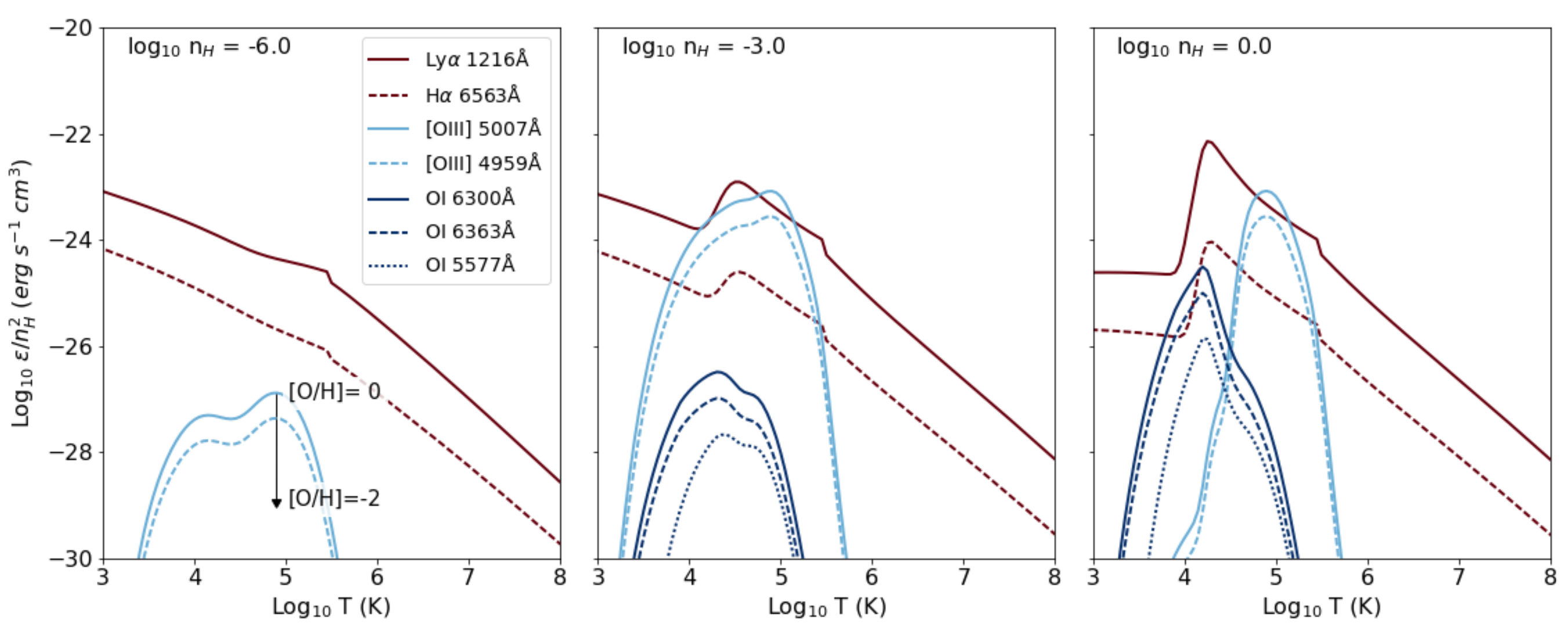} 
\caption{ 
The emissivity of strong hydrogen lines H$\alpha$ and Ly$\alpha$, as well as visible-wavelength oxygen lines, as a function of temperature for $z$ = 0, solar abundance and number densities $n_H$ = 1 cm$^{-3}$, 10$^{-3}$ cm$^{-3}$, and 10$^{-6}$ cm$^{-3}$ in the left, middle, and right panels, respectively.  Lines from the same ion are shown in the same color.  An arrow is drawn in the left plot indicating the vertical shift that would occur for all the oxygen line emissivities when scaling from solar abundance to 0.01 solar abundance.
}
\label{FigEmTable}
\end{figure*}

In Section 2, we briefly describe the EAGLE simulation and the numerical methods used to create emission maps.
In Section 3, we describe the results of the EAGLE simulation.
In Section 4, we apply the sensitivities of current instruments to the results from the simulation to determine the visibility of diffuse optical emission from the IGM and CGM.
Throughout this paper, we assume a standard $\Lambda$CDM cosmology with \citet{plan14} cosmological parameters: $\Omega_m$~=~0.307, $\Omega_\Lambda$ = 0.693, $\Omega_b$~=~0.048~25, $h = H_0/(100$~km~s$^{-1}$~Mpc$^{-1})$ = 0.6777.  It should also be noted that throughout this paper all box sizes (as well as particles masses and gravitational softening lengths) are \textit{not} quoted in units of $h^{-1}$.

\section{Numerical Methods}

\subsection{The EAGLE simulations}

The EAGLE suite \citep{scha15} is a set of cosmological, hydrodynamical simulations of the standard $\Lambda$ cold dark matter Universe where the values for cosmological parameters are taken from the 2014 Planck results \citep[as stated in the previous section;][]{plan14}.
The simulations are produced with a modified version of the \textit{N}-Body Tree-PM smoothed particle hydrodynamics (SPH) code \textsc{gadget 3} \citep{spri05}.  The subgrid physics are based on the prescriptions applied in the Over Whemingly Large Simulation (OWLS) project \citep{scha10}, which has been used previously to investigate UV and x-ray line emission via cooling channels of diffuse IGM gas \citep[e.g.][]{bert10a,bert10b,bert12,bert13,vand13}.  
The simulations include subgrid models for radiative cooling, star formation, stellar mass-loss and metal enrichment, energy feedback from star formation, gas accretion onto supermassive black holes, mergers of supermassive black holes and AGN feedback.  
Compared to OWLS, EAGLE has updated implementations of energy feedback from star formation \citep{dall12}, accretion of gas onto black holes \citep{rosa15,scha15}, and the star formation threshold \citep{scha04}.

In this study, we use the \textsc{reference} simulation at redshift $z$ = 0 with a box size of 100 comoving Mpc, which contains 1504$^3$ particles with initial gas particle masses of 1.81$\times10^6$~M$_{\odot}$ and dark matter particle masses of 9.70$\times10^6$~M$_{\odot}$. The comoving gravitational softening is set to 2.66 kpc, but is limited to 0.70 proper kpc from above.  The box size and resolution of this simulation are well-suited for studying the large scale structure while still resolving galaxies\footnote{In Appendix A1 we carry out a resolution test to determine the effects of increasing resolution on the results of this study.}.

The methods used to calculate the gas metal-line emission and create emission maps follow the prescriptions of \citet{bert10b}.  We refer the interested reader to that work for more details, while giving a brief outline of the procedure here.

We used the line emissivity tables created by \citet{bert10a}, which were also used in \citet{bert10b}, \citet{bert12}, \citet{bert13} and \citet{vand13}.
The gas emissivity tables for each line were computed as a function of temperature, density and redshift with \textsc{cloudy} version c07.02.02 {\citep{ferl98}}, under the assumptions of solar abundances, dust-free, optically thin gas and (photo)ionization equilibrium in the presence of the CMB and the \citet{haar01} model for the evolving UV/X-ray background radiation from galaxies and quasars.  Though more recent versions of \textsc{cloudy} exist, we use the same version of \textsc{cloudy} used by \citet{wier09} in order to ensure full self-consistency with the radiative cooling rates used in the EAGLE simulation.
 Following the prescription of \citet{bert13}, we adopt a solar abundance of Z$_\odot$ = 0.0127 corresponding to the default abundance set of \textsc{cloudy} version c07.02.02, which are a combination of abundances from \citet{alle01, alle02} and \citet{holw01} and may differ strongly from those estimated by \citet{lodd03}.  In particular, the oxygen abundance adopted here is about 20 percent smaller than that of \citet{lodd03}.  This should be kept in mind when comparing results of different studies, but we stress that the assumed solar abundances play no role when computed the emission from the EAGLE simulation, which is calculated using the absolute abundance predicted by the simulation.
The tables include a total of about 2000 emission lines for 11 elements.  The temperature is sampled in bins of $\Delta$log$_{10}T$ = 0.05 in the range 10$^2$ $<T<$ 10$^{8.5}$ K and the hydrogen number density in bins of $\Delta$log$_{10}n_H$~=~0.2 in the range 10$^{-8}$ $< n_H$ $<$ 10 cm$^{-3}$.  
Plots based on the emissivity tables are shown in Fig.~\ref{FigEmTable} for select hydrogen and oxygen lines at z = 0 and solar abundances, with three hydrogen number densities: $n_H$~=~1~cm$^{-3}$,~10$^{-3}$~cm$^{-3}$,~and~10$^{-6}$~cm$^{-3}$.  
Visible line transitions for hydrogen and oxygen are included, as well as Ly$\alpha$ for reference.  
Note that when creating the emission maps, the emissivities are scaled by the ratio of the particle abundance to solar abundance (as described in Section 2.2).  We demonstrate this scaling with an arrow in the left plot of Fig.~\ref{FigEmTable}, which indicates the decrease in emissivity  between solar abundance and 0.01 solar abundance (i.e. a downwards vertical shift of the solar abundance emissivity curves in log-space).

The assumption of negligible self-shielding (i.e.\ optically thin gas) may break down in high density \citep[i.e.\ n$_H \gtrsim 10^{-2}$ cm$^{-3}$;][]{rahm13}, low temperature (i.e.\ T~$\lesssim10^4$~K) cases.  
At low densities, the hydrogen gas is predominantly photo-ionized (indicated by the smooth curves of the n$_H$ $=$ 10$^{-6}$ cm$^{-3}$ plot of Fig.1), but the gas is too diffuse for self-shielding to become important.  
In the dense gas, collisional excitation dominates at temperatures T~$>$~10$^4$~K (indicated by the sharply peaked curves for hydrogen lines in the n$_H$~=~1~cm$^{-3}$ plot of Fig.1) and produces the brightest oxygen and hydrogen line emission, but a significant fraction of the hydrogen emissivity is also emitted from gas with T~$<$~10$^4$~K, which is photo-ionized (the tail of the hydrogen line curves at low temperature in the right panel of Fig.1).  
In addition, at transitional densities between these two regimes of ionization (e.g. middle panel of Fig.1) the emissivity from photo-ionization at T~$<$~10$^4$~K for the hydrogen lines is comparable in strength to the emissivity from collisional ionization that peaks at T~$>$~10$^4$~K.  For T~$\lesssim$~10$^4$~K and n$_H$~$\gtrsim$~10$^{-2}$~cm$^{-3}$ \citep[e.g.][]{rahm13a}.  Self-shielding may then expected to be important for the hydrogen line emission, while the [OIII] oxygen line emission, which peaks at T~$>$~10$^4$~K, may be less affected by self-shielding.
At the densities and temperatures where self-shielding is expected to be important, though, the radiation from stellar sources, which is not included, is expected to become just as important as the UVB and may counteract the effects of self-shielding \citep[e.g.][]{rahm13a}.   
We break down the contribution to the H$\alpha$ emission from different sources, including star-forming and self-shielded gas, in Appendix A2.

The assumption of ionization equilibrium (both for the cooling rates and the emissivity tables) is justified for regions where they are predominantly photo-ionized, but in the WHIM and the outer regions of clusters, non-equilibrium processes may become important for metals \citep[e.g.][and references therein]{gnat09,bert10b,oppe13}. 
\\

\subsection{The emission maps}

\begin{figure*}
\centering
\includegraphics[width=0.99\linewidth]{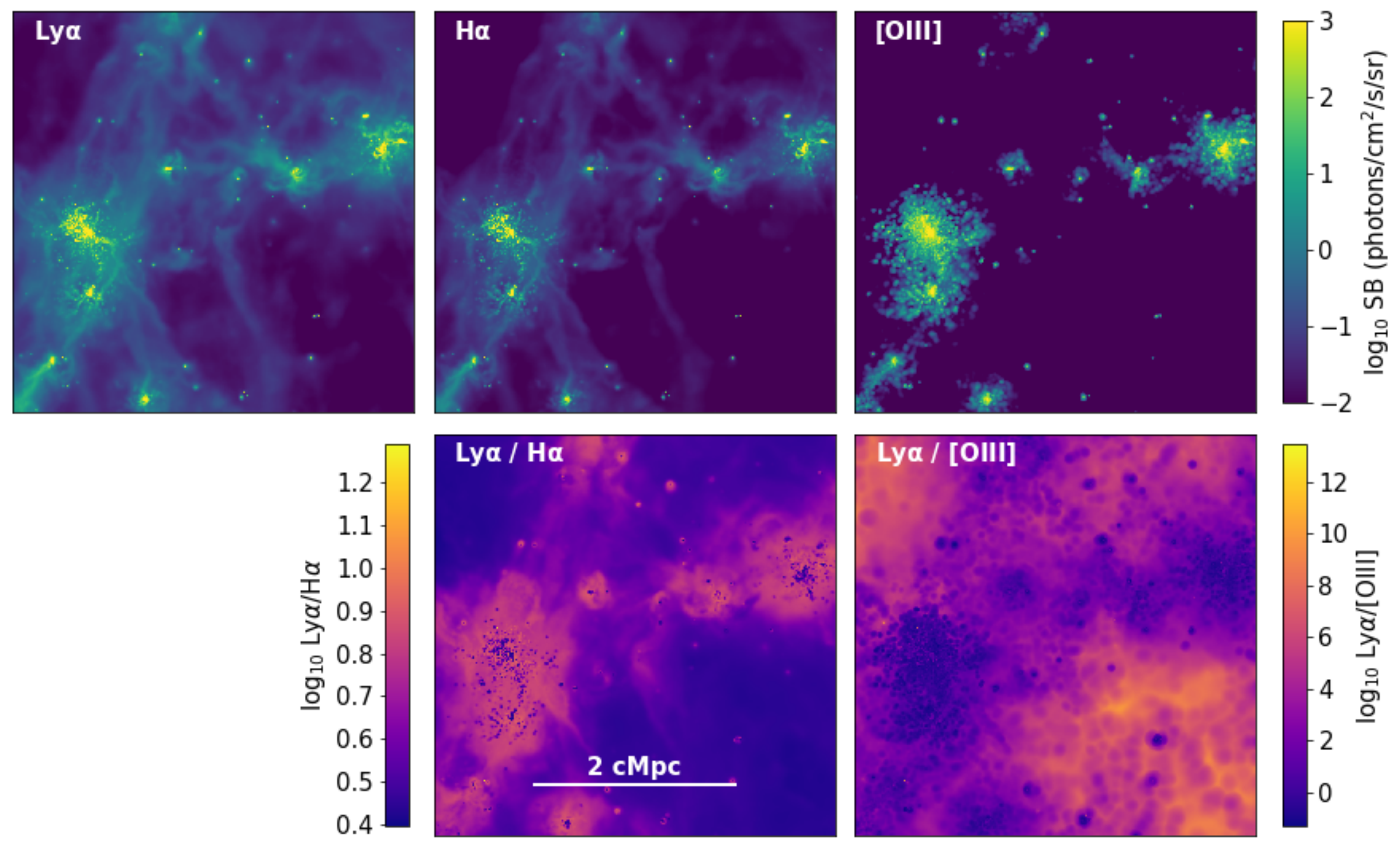} 
\caption{ Surface brightness maps of line emission at redshift $z$ = 0 projected from EAGLE non-star-forming particles for the line transitions H$\alpha$, Ly$\alpha$, and [OIII] 5007 $\mathrm{\AA}$, in the top-left, top-middle, and top-right panels, respectively.  Each map is 4$\times$4 comoving Mpc in size.  Also shown are ratio maps of Ly$\alpha$ to H$\alpha$ emission and Ly$\alpha$ to [OIII] 5007 $\mathrm{\AA}$ emission, in the bottom-middle and bottom-right panels, respectively.  Note that the surface brightness scale is the same for the top row of panels, but is different for each of the ratio maps.
}
\label{HalphaLyalphacomparison}
\end{figure*}

The procedure for computing the surface brightness emission maps follows that used in \citet{bert10b}, though we will include a brief description here for reference.

In OWLS, a constant threshold of n$_H$~$>$~10$^{-1}$~cm$^{-3}$ was used to delineate when gas would become star forming: above this density a cold phase is expected to form \citep{scha04}.  EAGLE instead uses a metallicity-dependant density threshold, which takes into account the fact that the transition between the warm neutral phase and the cold molecular phase occurs at lower density and pressure if the metallicity is higher \citep{scha04,scha15}.  Since EAGLE does not model the cold gas phase \citep[instead imposing a temperature floor according to a polytropic equation of state;][]{scha08}, we either set the emission from star-forming gas to zero or use an empirically motivated prescription to calculate H$\alpha$ emission from the star-forming gas.  The empirical prescription takes the rate of star formation in the gas and the measured conversion factor, $C_x$, between star formation rate and intrinsic H$\alpha$ luminosity \citep[specifically, log~$\dot{M}_{\star}$~=~log~$L_x$~--~log~$C_x$, where log ($C_x$ / erg s$^{-1}$ M$_{\odot}^{-1}$ year) = 41.27;][]{kenn12} to calculate the amount of H$\alpha$ emission from the star-forming gas.  In other words, for star-forming gas, we are assuming that the emission is dominated by recombination radiation from HII regions\footnote{Note that we neglect dust extinction, which is typically between 0 and 1 mag at H$\alpha$ \citep{kenn12}.}.
We assign star-forming emission based on this empirical calibration of the star-forming gas rather than modelling star particles as single stellar populations to estimate the H$\alpha$ emission from star-forming regions due to the low number of young star particles in the simulation which would cause poor sampling.
Note that resonant scattering is neglected, but is expected to be important for the distribution of Ly$\alpha$ emission \citep[e.g.][]{bert12,fauc10}.

To estimate the surface brightness in emission lines from the simulation to use to predict the detectability of the diffuse emission, the properties of the particles in the simulation box are projected onto a spatial grid, then slices of this projected box are taken in radial distance. Specifically, the luminosity, $L$, of the particle is $L_{y,i}~=~\epsilon_{y,\odot}$($\rho_i$,T$_i$,z)$~V_i~(X_{y,i}/X_{y,\odot})$~erg~s$^{-1}$, where the element is designated by $y$, the particle identifier is $i$, $\epsilon(\rho,T,z)$ is emissivity interpolated from the \textsc{cloudy} tables at solar abundance, $V$ is the volume of the particle calculated from the particle mass and density, and $X$ is the mass fraction, using SPH-smoothed abundances.  Explicitly, $X_{y,i}$ is the mass fraction of element $y$ in particle $i$, and $X_{y,\odot}$ is the solar mass fraction of element $y$.  We note again that we omit star-forming gas when calculating the emission using the \textsc{cloudy} tables, so there is no double accounting for the emission from the star-forming regions.
The flux from the particle is 
\begin{equation}
F_{i} = \frac{L_i}{4\pi D_{L}^{2}} \frac{\lambda(1 + z)}{h_P c}
\end{equation}
in units of photons cm$^{-2}$ s$^{-1}$, where $h_P$ is Planck's constant, $D_L$ is the luminosity distance of the emitter, $\lambda$ is the rest-frame wavelength of the emitted photons and $c$ is the speed of light.  
The fluxes from each particle are projected onto a 2D grid, then
the surface brightness is found by dividing the flux by the solid angle, $\Omega$, subtended by a pixel of the 2D grid, i.e. SB = F/$\Omega$.

For our analysis, we use emission maps from the 100 Mpc box simulation, with a slice width of 20 comoving Mpc. The depth of the slice, 20 comoving Mpc, corresponds to a wavelength shift of $\approx$ 3 nm or $\approx$ 1400 km/s at $\lambda$ = 656.3 nm (of order the average velocity dispersion of galaxy clusters).

Emission maps for Ly$\alpha$, H$\alpha$, and [OIII] 5007 $\mathrm{\AA}$ are shown in the top row of panels in Fig.~\ref{HalphaLyalphacomparison} encompassing a node of the cosmic web where a galaxy group has formed.  These maps are created from the simulation at redshift $z$~=~0, and are 4$\times$4 comoving Mpc on a side.  The physical resolution of the emission maps is 6.25 kpc per pixel (for reference, this corresponds to an angular resolution of $\approx$~10 arcsec for structures at a radial distance of 75 Mpc, while the total length, 4 comoving Mpc, corresponds to an angular scale of $\approx$~1.8$^\circ$).
Only non-star-forming particles are included in the emission maps of Fig.~\ref{HalphaLyalphacomparison}.
In Fig.~\ref{HalphaLyalphacomparison}, we also show maps of the ratio of emission between Ly$\alpha$ and H$\alpha$ (bottom-middle panel) and between Ly$\alpha$ and [OIII] 5007 $\mathrm{\AA}$ (bottom-right panel)\footnote{We note that these ratio maps are ratios of emission in photons: to convert to ratios of emission in energy, one can simply multiply by the ratio of the line wavelengths.}.  
The Ly$\alpha$ and H$\alpha$ emission trace the diffuse gas in the simulation, whereas the oxygen line emission is concentrated in the denser gas pockets.  
Though Ly$\alpha$ and H$\alpha$ emission are produced by similar mechanisms -- predominantly photo-ionization that increases in strength as temperature decreases -- 
the Ly$\alpha$ to H$\alpha$ ratio is not constant due to the presence of different sources of emission, which include collisional excitation, collisional ionization, and photoionization.  At different temperatures and densities, different emission sources become significant, which produces various ratios of Ly$\alpha$ to H$\alpha$ photons.  
In practice, the Ly$\alpha$ emission is brighter by up to a factor of $\approx$ 20 in emission compared to the H$\alpha$ emission, but for the majority of the diffuse emission, the relative surface brightness of Ly$\alpha$ to H$\alpha$ is $\approx$~8. 

It is interesting to note that the oxygen line emission is relatively strong -- stronger than both the Ly$\alpha$ and the H$\alpha$ emission -- in dense pockets of gas, where the [OIII] 5007 $\mathrm{\AA}$ emission is brighter than the Ly$\alpha$ emission by up to an order of magnitude.  This contrasts with emission from diffuse structures, where the Ly$\alpha$ emission dominates by many orders of magnitude.
Though the [OIII] lines have strongly peaked emission at the temperatures of the WHIM (as seen in Fig.~\ref{FigEmTable}), it is predominantly collisionally ionized and the strength of the emission also depends on the abundance of the ion and the density of the gas, which boosts the oxygen emission in dense pockets where the metallicity is higher rather than in the diffuse cosmic web. 

It would be valuable to measure the oxygen emission to place constraints on the metallicity and exchange of material from the galaxies to the CGM and IGM.  From this simple comparison, it appears that the oxygen emission will have similar detectability to the H$\alpha$ that we find here (if not being more detectable). While the following analysis and discussion focuses on H$\alpha$ emission, our findings for the detectability of H$\alpha$ emission from the CGM can be applied to [OIII] emission, as well.
Finally, we re-iterate that we ignored emission from the interstellar medium (i.e. the star-forming gas), which may dominate the brightest regions.

\section{Instruments}


\begin{deluxetable*}{c c c c c c c}
\tabletypesize{\scriptsize}
\tablecaption{The characteristics of KCWI\tablenotemark{a} \citep{morr12}, PCWI \citep{matu10}, MUSE \citep{baco10}, and FIREBall \citep{quir14}, as well as the redshift range for which the Ly$\alpha$ and H$\alpha$ transitions fall into the wavelength range of the instrument.}
\label{CharTable}      
\tablewidth{0pt}
\tablehead{
\colhead{Instrument} & \colhead{Wavelength range} & \colhead{FOV} & \colhead{Pixel Size} & \colhead{$z$ range} & \colhead{$z$ range} & \colhead{Spectral Resolution}\\ 
 \colhead{} & \colhead{($\mathrm{\AA}$)} & \colhead{(arcsec)} & \colhead{(arcsec)} & \colhead{Ly$\alpha$ (1216$\mathrm{\AA}$) } &  \colhead{H$\alpha$ (6563$\mathrm{\AA}$) } & \colhead{}
}
\startdata
     
     PCWI         &      3800 -- 9500   &     60 $\times$ 40                &      2.5 $\times$ 1                   &      2.1 -- 6.8     &       0 -- 0.45   &  5000 \\
     KCWI         &      3500 -- 10500 &     20 $\times$ (8 to 33)      &      0.5  $\times$ (0.35 to 1.4) &      1.9 -- 7.6     &       0 -- 0.60   &  900 to 18000\tablenotemark{b} \\
     MUSE        &      4650 -- 9300   &     60 $\times$ 60                &      0.2                                     &      2.8 -- 6.6     &       0 -- 0.42   & 1750 -- 3750\tablenotemark{c} \\
     FIREBall-2  &     2000 -- 2080    &    1200 $\times$ 1200        &      4                                        &      0.64 -- 0.71 &       --               & 2150
\\
\enddata

\tablenotetext{a}{Note that the full proposed KCWI wavelength coverage is listed here. KCWI currently has only a blue channel with wavelength range of 3500 -- 5600 $\mathrm{\AA}$, which does not cover the H$\alpha$ transition.}
\tablenotetext{b}{Depends on chosen grating and IFU slicer configuration.}
\tablenotetext{c}{Smoothly varies from the blue end to the red end of the wavelength range.}

\end{deluxetable*}

We now turn to the practical aspects of the detectability of emission from the IGM and CGM, starting with a consideration of instruments.
There are a number of ground-based instruments that have come online in the last few years that are designed to probe diffuse emission from the IGM at z~$>$~1.5 and may also be applied to imaging the cosmic web through visible wavelength emission.  These include the Cosmic Web Imager \citep[PCWI;][]{matu10} on the 200" Hale Telescope at Palomar, the Multi-Unit Spectroscopic Explorer \citep[MUSE;][]{baco10} on the Very Large Telescope (VLT), and the Keck Cosmic Web Imager \citep[KCWI;][]{morr12}.  We note that the balloon-borne experiment FIREBall \citep{quir14} is designed to image Ly$\alpha$ emission from the cosmic web at redshift $z$ $\sim$ 0.7, but its wavelength range does not include visible wavelength emission.  Another note is that the KCWI wavelength range currently does not cover the H$\alpha$ line (the blue channel covers 350 nm -- 560 nm), but the planned red channel will open the full wavelength range to 350 nm -- 1050 nm and allow H$\alpha$ studies.
PCWI, KCWI, and MUSE have wavelength ranges in the visible spectrum and have reached extremely low surface brightness limits targeting low surface brightness emission from the circumgalactic and intergalactic medium at high redshift\footnote{While it is difficult to compare the surface brightness limits reached by instruments due to differences in observing conditions and modes, we note that in observations targeting low surface brightness extended emission (over a 10 -- 15 arcsec scale), PCWI has reached a detection limit of $\sigma_{SB}\approx~1.3\times10^{-19}$ erg s$^{-1}$ cm$^{-2}$ arcsec$^{-2}$ \citep{mart14} and MUSE reaches the same depth for an aperture of 1 arcsec \citep{wiso18}.}.
Here we focus on the detectability of H$\alpha$ emission with Dragonfly, but the characteristics of these instruments are listed in Table 1, for reference.

Dragonfly is currently being upgraded to support narrow-band imaging work.  
In its present 48-lens configuration, the telescope is equivalent to a 0.99 m aperture, f/0.4 telescope, with a 2.6$^\circ \times 1.9^\circ$ field-of-view.   Dragonfly's large field-of-view and low resolution combined with its low surface brightness capabilities make it uniquely suited to imaging spatially very extended, extremely low surface brightness structures, such as ultra diffuse galaxies \citep{abra14}.
Dragonfly has imaged down to surface brightnesses around $\sim$32 mag arcsec$^{-2}$ in $g$-band.
To determine the sensitivity of Dragonfly as a narrow band imager, we will use the following specifications to describe the telescope system: the transmittance of the lenses ($\tau_l$ = 0.85) and filters ($\tau_f$ = 0.95), a narrowband filter width of 3 nm, the quantum efficiency of the detectors ($QE = 0.70$), along with their dark current ($D = 0.04$ electrons s$^{-1}$ pixel$^{-1}$) and read noise ($R = 2$ electrons pixel$^{-1}$).

An estimate of the sky background within the filter bandwidth is found by integrating the flux of the Gemini model spectrum of the sky background within the bandwidth of the Dragonfly narrow band filters.
In this case we take a realistic assumption for the Dragonfly Telescope observing conditions, where on average 50\% of nights are darker than the adopted sky brightness.  This value is obtained from a Gemini model sky spectrum\footnote{\texttt{http://www.gemini.edu/sciops/ObsProcess/obsConstraints/ atm-models/skybg$\_$50$\_$10.dat}}, which is scaled to match the sky brightness at 50\%-ile (at around $\lambda$ = 656.3 nm the integrated sky background within the filter width is $\approx$2.2$\times 10^{6}$~photons ~s$^{-1}$~nm$^{-1}$~cm$^{-2}$~sr$^{-1}$)\footnote{The sky background continuum in between sky lines has not been measured at the location of the Dragonfly Telescope, but sky background measurements are found to be roughly consistent between different locations \citep[e.g ][]{hanu03,benn98} and the Gemini model is consistent with the sky background measured by the Ultraviolet and Visual Echelle Spectrograph on VLT \citep{hanu03}.
The sky brightness varies daily and depends on a series of factors.  The most important of these are the lunar phase and the target-moon separation, but other factors include the ecliptic latitude, zenith angle, and the phase of the solar cycle.  These factors cause the sky brightness to vary from fractions of a magnitude \citep[for ecliptic latitude, solar maximum and airmass up to 1.5; e.g.][]{benn98} up to $\approx$4 magnitudes \citep[due to lunar phase and the target-moon separation; e.g.][]{kris91}.  Increases in sky brightness of more than $\approx$1 magnitude can be avoided by not observing when the moon is up or when the moon is closer than 30$^\circ$ to the target.  This is the largest source of uncertainty in low surface brightness measurements and requires careful monitoring of the sky background signal to properly subtract and account for.}.

\begin{figure*}
\centering
\includegraphics[width=0.99\linewidth]{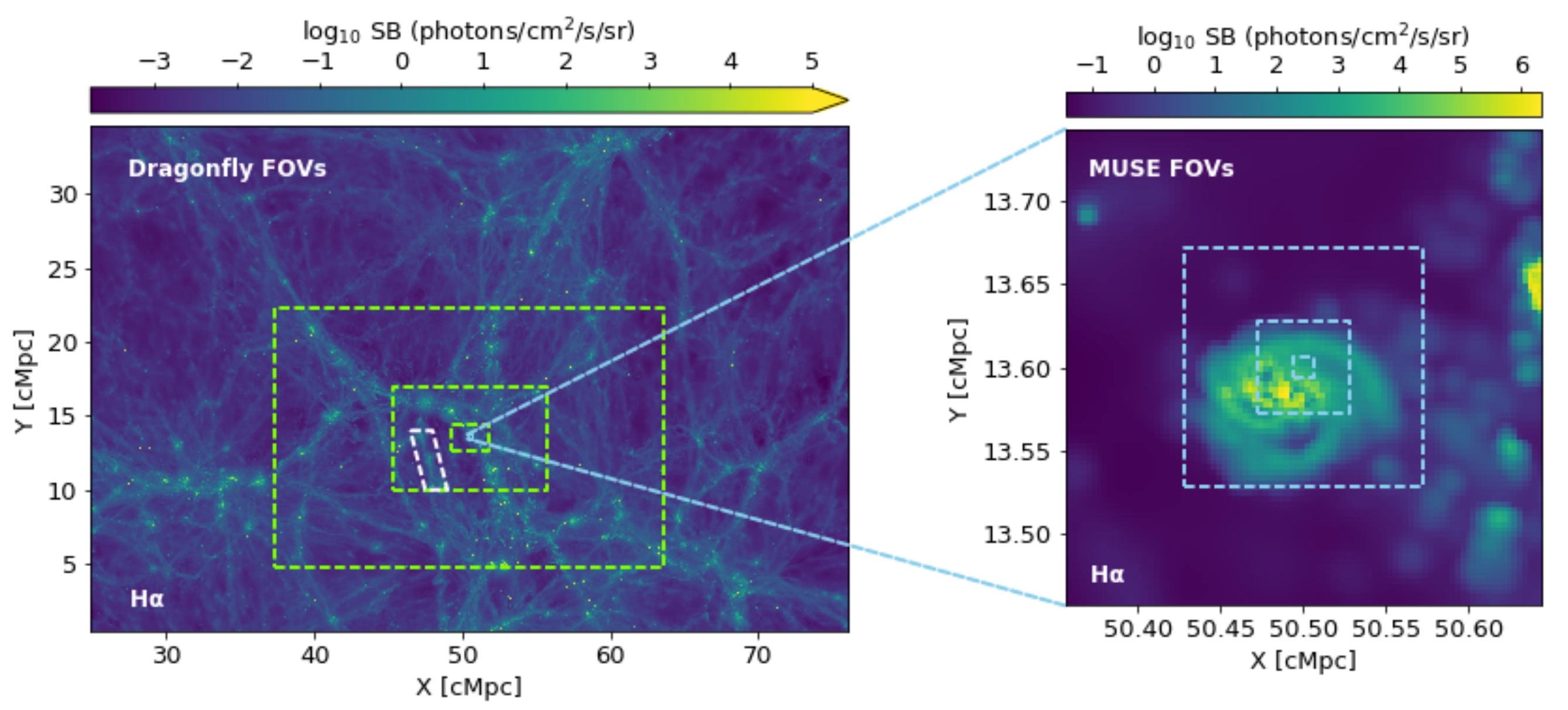}
\caption{  H$\alpha$ surface brightness mapped from the full EAGLE 100 Mpc simulation box at redshift $z$ = 0 (with a 20 Mpc width) projected to the size of the Dragonfly field-of-view, 2.6$^\circ \times 1.9^\circ$, (left panel) and the size of the MUSE field-of-view, 60" $\times$ 60", (right panel) at redshift $z$ $\sim$ 0.24 (corresponding to a radial distance of $\sim$1000 Mpc).  The dashed green (blue) lines in the left (right) panel correspond to the size of the Dragonfly (MUSE) field-of-view at redshifts of $\sim$ 0.01, 0.05, and 0.12 or radial distances of 50 Mpc, 200 Mpc, and 500 Mpc.  An example filament of the IGM is indicated by the white dashed box.
}
\label{FigMaps}
\end{figure*}

With these values, the signal-to-noise ratio (SNR) can be calculated as:
\begin{equation}
SNR = \frac{I t}{\sqrt{(I + B n + D n) t + R^2 n}},
\end{equation} 
where $I$ is the count rate, $B$ is the sky background per pixel, and $n$ is the number of pixels.  The exposure time, $t$, is usually given in seconds.  Both $I$ and $B$ depend on the total transmittance of the camera, $\tau = \tau_l\times\tau_f$.

Equation (2) indicates that with a 3 nm narrowband filter on Dragonfly, a surface brightness of 1000 photons  s$^{-1}$ cm$^{-2}$ sr$^{-1}$ can be reached with a signal-to-noise ratio $\approx$~5 in $\approx$~60 hours when targeting 100 arcsec features (see Fig.~\ref{FigSNR}).  As we will now show, the structures in the local Universe are very large. By exploiting its large field of view, Dragonfly is likely to be able to probe the IGM and CGM in the local Universe down to depths similar to those reached by KWCI and MUSE on much larger telescopes at high redshifts.

The spatially resolved spectrometers mentioned above were designed to image the high redshift cosmic web with their relatively small fields of view matched to the angular scale of the cosmic web at redshift $z$ $>$ 1.5.  The field-of-views for each instrument are maximally 60$\times$40 arcsec$^2$ for CWI, 20$\times$34 arcsec$^2$ for KCWI, and 60$\times$60 arcsec$^2$ for MUSE. 
In Fig.~\ref{FigMaps}, we compare the MUSE field-of-view to the Dragonfly field-of-view by projecting the fields-of-view onto the EAGLE simulation.  The dashed lines outline the size of the Dragonfly/MUSE field-of-view when targeting structures in the local Universe at distances of $\approx$ 50 Mpc, 200 Mpc, and 500 Mpc, and 1000 Mpc (corresponding to redshifts of $z$ $\approx$ 0.01, 0.05, 0.12, and 0.24).  
Fig.~\ref{FigMaps} demonstrates that it may be possible for the spatially resolved spectrometers to observe H$\alpha$ emission from the CGM of local galaxies, while the filamentary structures of the IGM in the local Universe extend to far larger scales than their fields of view.  
An additional consideration is the effect of scattering in the telescope optics:  in typical telescope optical design, the scattering of light from central star-forming regions causes the surface brightness background level to rise and may wipe out the signal from the extremely faint diffuse gas.
Dragonfly's all-refractive design minimizes scattered light, so, in principle, it is particularly well-suited to probing the local CGM and IGM.  We explore this idea further in the next section.

\section{Detectability of the CGM and IGM in the local Universe}
\subsection{CGM}

\begin{figure*}
\centering
\includegraphics[width=0.99\linewidth]{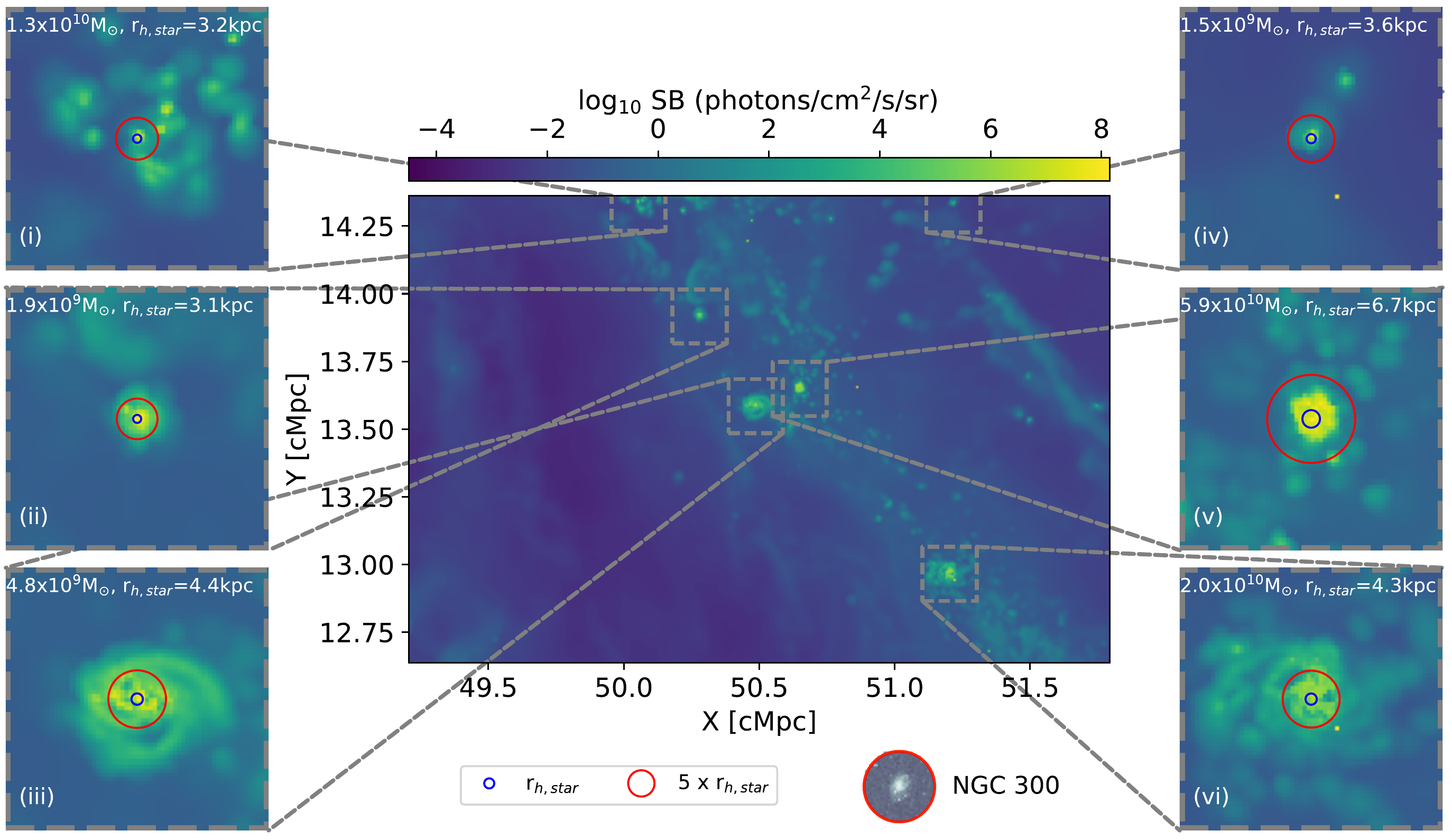}
\caption{ 
The central panel depicts a cutout from the EAGLE simulation in H$\alpha$ emission that is the size of the Dragonfly field-of-view at a distance of 50 Mpc, with a slice thickness of 20 comoving Mpc.  The physical resolution in each map is 3.125 kpc per pixel, which corresponds to $\approx$ 13 arcsec angular resolution.  Zoom-ins on galaxies from the cutout that have stellar masses greater than 10$^9$ M$_{\odot}$ are shown on either side (each zoom-in is 200 comoving kpc on a side).  The stellar masses and half-stellar mass radii ($r_{h, star}$) of the selected galaxies are labelled on each zoom-in.
The blue circles overplotted on the zoom-ins correspond to $r_{h, star}$ of each galaxy.
The red circles overplotted on the zoom-ins correspond to 5$\times r_{h, star}$ for the selected galaxy -- generally the limit for detections of stars in galaxies \citep[surface brightnesses of less than 32 mag/arcsec$^2$; ][]{zhan18,blan05}.  
An inset image of NGC300 from the DSS is provided for reference, with the outermost radius corresponding to the red circle.
Note how coherent structures are traced by diffuse line emission that extends far beyond the stellar components of the galaxies.
}
\label{FigFOVwithgalcutouts}
\end{figure*}

\begin{figure*}
\centering
\includegraphics[width=1.00\linewidth]{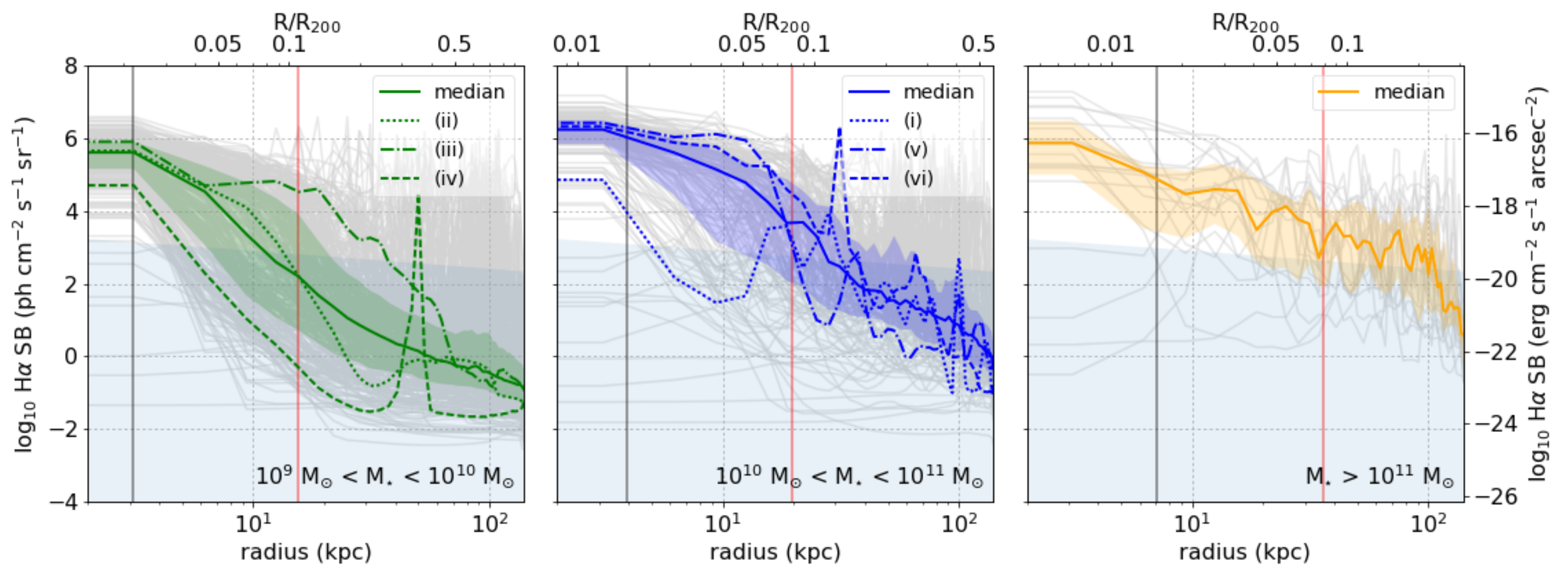}
\caption{ Azimuthally averaged radial H$\alpha$ surface brightness profiles of galaxies from a 20 comoving Mpc slice through the EAGLE simulation at redshift $z$ = 0 (pictured in the left panel of Fig.~\ref{FigMaps}).  The radial profiles of galaxies with stellar masses of 10$^9$ M$_{\odot}$ -- 10$^{10}$ M$_{\odot}$ (10$^{10}$ M$_{\odot}$ -- 10$^{11}$ M$_{\odot}$; 10$^{11}$ M$_{\odot}$ and up) are shown in light grey in the left (middle; right) plot with their median plotted in green (blue; orange) and the same-colored shading filling the area between the 25th and 75th percentiles.  The half-stellar mass radii ($r_{h, star}$)  and 5$\times r_{h, star}$ of the galaxies are indicated with black and red vertical lines, respectively.  
The 3$\sigma$ limit for detection with Dragonfly in a 100 hour exposure is indicated by edge of the light blue shaded area (i.e. above the shaded area, detections would be made with $>$3$\sigma$ confidence).
For the top x-axis of each panel, the radius has been divided by the median R$_{200}$, the radius within which the mean internal density is 200 times the critical density, $3H^2/8\pi G$, centred on the dark matter particle of the corresponding halo with the minimum gravitational potential \citep{scha15}.
The individual surface brightness profiles of the six galaxies identified from the Dragonfly field-of-view in Fig.~\ref{FigFOVwithgalcutouts} are also plotted in color on the panels with the mass range corresponding to the stellar masses of the individual galaxies.   The individual galaxies are labelled with numbers corresponding to the same objects in Fig.~\ref{FigFOVwithgalcutouts}.  Of the six galaxies, those with masses between 10$^9$ M$_{\odot}$ and 10$^{10}$ M$_{\odot}$ are plotted in green, and the galaxies with stellar mass between 10$^{10}$ M$_{\odot}$ and 10$^{11}$ M$_{\odot}$ are plotted in blue, with varying linestyles to differentiate between the individual galaxies. }
\label{FigHalphaProfile}
\end{figure*}

In this Section, we move from the general considerations of the previous section to explore predictions for the visibility of the local CGM and IGM in detail.  For the following analysis, we will specifically consider H$\alpha$ emission because it closely traces the gas in the diffuse CGM and IGM and is accessible from the ground.
We assume Dragonfly with 3 nm bandwidth filters mounted at the entrance apertures of each lens in the array (the configuration is described in detail in Lokhorst et al.\ in preparation, an instrumental companion to the present paper).
We include emission from both star-forming and non-star-forming particles for the following analysis (see Section 2.2 for details) and use the EAGLE Galaxy Catalogue \citep{mcal16} to select galaxies by stellar masses, half-stellar mass radii, half-gas mass radii, and location.

In Fig.~\ref{FigFOVwithgalcutouts}, the Dragonfly field-of-view when imaging structures 50~Mpc away is shown centered on a sample region from the EAGLE simulation.  
The slice thickness of the simulation is the same as that used in Section 2.2 (i.e. 20 comoving Mpc), where the entire slice is assumed to be at the same redshift.
The field is centered on a typical filament of the cosmic web, with boxes drawn around all galaxies with stellar masses greater than 10$^9$~M$_{\odot}$ within the field-of-view.  Zoom-ins for each galaxy are also shown where each cutout has side lengths of 200~kpc.  From the zoom-ins, it is clear that the galaxies in the EAGLE simulation have a wide variation of gas properties, both in their mass and distribution.  On each of the zoom-ins, the blue circles are drawn at the half-(stellar mass) radius ($r_{h, star}$) of each galaxy, and red circles indicate 5$\times r_{h, star}$ for the galaxy.  The limit of 5$\times r_{h, star}$ corresponds to the radial limit for detections of stars in galaxies when imaging down to surface brightnesses fainter than 32 mag/arcsec$^2$ \citep{zhan18,blan05}.  For reference, an inset of NGC 300 is shown where its outermost radius corresponds to 5$\times r_{h, light}$.  Note how coherent structures are traced by diffuse line emission that extends far beyond the stellar components of the galaxies.

Azimuthally-averaged radial H$\alpha$ surface brightness profiles around galaxies in the EAGLE simulation are shown in Fig.~\ref{FigHalphaProfile}.
The median radial profiles for galaxies within a specified mass range are shown in Fig.~\ref{FigHalphaProfile}, superimposed upon the backdrop of the individual profiles for each individual galaxy in light grey.  In the left (center; right) panel, all galaxies with stellar masses of 10$^9 M_{\odot}$--10$^{10} M_{\odot}$ (10$^{10} M_{\odot}$--10$^{11} M_{\odot}$; 10$^{11} M_{\odot}$ and up) are shown and the median profile is plotted in green (blue; orange), with a lighter-colored shaded area indicating the 25$^{th}$ to 75$^{th}$ percentiles.  The median virial radius for galaxies within each mass bin is indicated on the top x-axis (using the $R_{200}$ definition of the virial radius to normalize).
The individual radial H$\alpha$ profiles for the six galaxies with zoom-ins in Fig.~\ref{FigFOVwithgalcutouts} are also plotted in Fig.~\ref{FigHalphaProfile}: the three galaxies with stellar mass between 10$^9 M_{\odot}$ and 10$^{10} M_{\odot}$ are plotted in the left panel, and the three galaxies with stellar mass between 10$^{10} M_{\odot}$ and 10$^{11} M_{\odot}$ are plotted in the middle panel.  
The profiles in each mass bin are close to power-law in shape.
Note that galaxy (iii) is fainter than the more massive galaxies overall, but for some radii it has comparable H$\alpha$ brightness.
This is interesting because it implies that similarly bright extended halos can be found around a large mass range of galaxies, despite the marked difference when considering the statistical trends.  In addition, this demonstrates that though the average surface brightness profiles are useful for getting an idea of the brightness profile, individual profiles can be much brighter (or fainter) than the averages, making them much easier (or harder) to detect.
In the following sections we investigate the visibility of gas in the CGM, considering extended halos in Section 4.1.1, gas streaming into and around galaxies in Section 4.1.2, and photoluminescence from the cosmic ultraviolet background in Section 4.1.3.

\subsubsection{Predicted Visibility of Extended Halos}

Figures~\ref{FigFOVwithgalcutouts} and \ref{FigHalphaProfile} illustrate a predicted ``glow'' from the gas filling the halos of galaxies, which has not yet been observed locally. 
At higher redshifts this phenomenon was first detected by \citet{stei11}, who used deep narrowband imaging around the Ly$\alpha$ line to look for extended structure around very actively star-forming Lyman break galaxies at redshift $z$ $\sim$ 2.5.  By stacking 92 individual galaxies and azimuthally averaging, they found that there was an excess diffuse Ly$\alpha$ component that extended out to $\approx$~80 kpc (reaching surface brightness SB$_{Ly\alpha}\sim10^{-19}$ ergs s$^{-1}$ arcsec$^{-2}$ cm$^{-2}$), compared to the continuum emission, which stopped at $\approx$ 10 kpc.  Similar stacking analyses of thousands of star-forming galaxies in various environments at redshifts $z$ $\sim$ 3 -- 6 have followed which corroborate the existence of extended Ly$\alpha$ halos with luminosities and sizes that vary depending on the environment \citep[with various filtering and averaging methods these studies reach surface brightnesses SB$_{Ly\alpha}\sim10^{-19}-10^{-21}$ ergs s$^{-1}$ arcsec$^{-2}$ cm$^{-2}$ and radii $\approx$ 30 -- 80 kpc; e.g.][]{mats12,momo14,momo16, wiso18}.
With MUSE, \citet{wiso16,wiso18} detected halos of extended Ly$\alpha$ emission around \emph{individual} galaxies at redshift $\sim$ 3 -- 6, which extend out to $\approx$ 30 -- 70 kpc and reach surface brightnesses SB$_{Ly\alpha}\sim10^{-19}-10^{-20}$ ergs s$^{-1}$ arcsec$^{-2}$ cm$^{-2}$ through azimuthal averaging.  
In addition, extended Ly$\alpha$ nebulae around quasars at redshift $z$ $\sim$ 2 -- 4 have been observed to have sizes as large as $\sim$300 kpc at similar surface brightnesses \citep[e.g.][]{arri18,bori16,cai18}.

\begin{figure*}
\centering
\includegraphics[width=.95\linewidth]{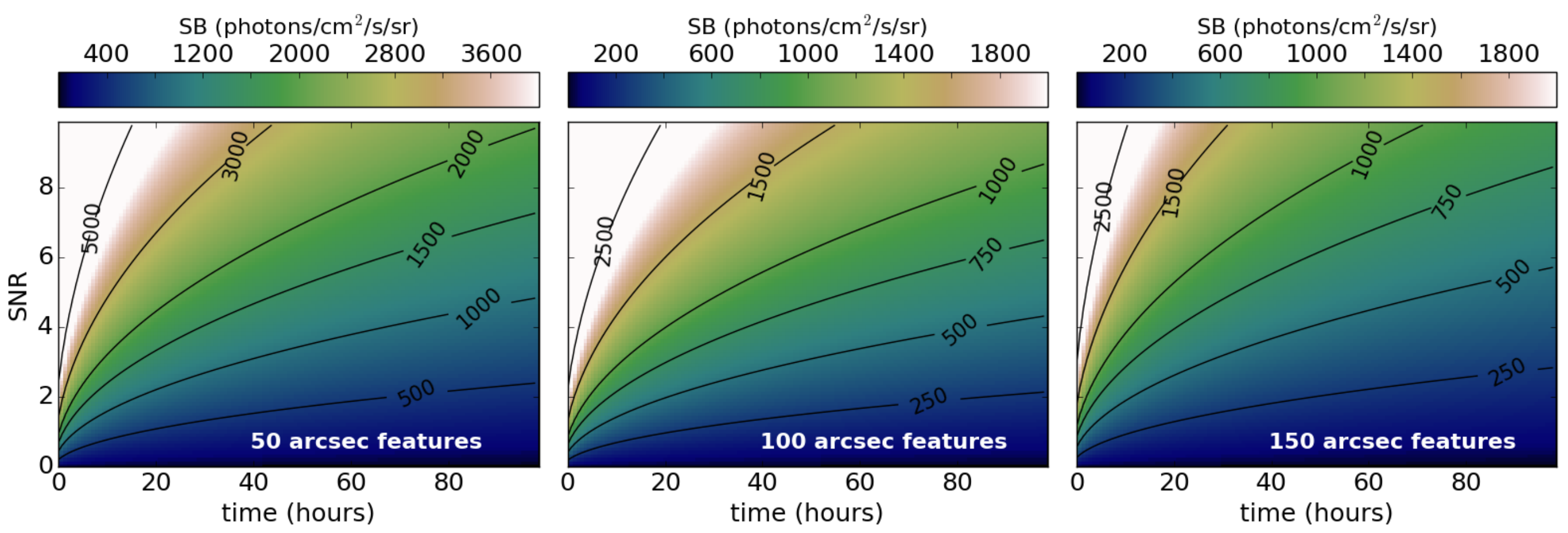}
\caption{
The signal-to-noise ratio for the Dragonfly Telescope narrow-band imaging at $\lambda$ = 656.3 nm as a function of integration time for specific surface brightnesses (indicated by the color map).  
The surface brightnesses are calculated for 50" (100"; 150") square features in the left (middle; right) plot.  In addition to the color map, the contours also show surface brightness to guide the eye.
}
\label{FigSNR}
\end{figure*}

We can use these existing high-redshift results to check the reasonableness of the numerical simulations we have just shown.
To compare our predictions with the higher redshift observations, we estimate the strength of H$\alpha$ emission in the extended halo by converting the Ly$\alpha$ surface brightness measurements at higher redshift to H$\alpha$ estimates through a series of physical relations.  For this estimate, we use the \citet{stei11} results, which are fairly representative of the various high redshift Ly$\alpha$ emission measurements, and would correspond to highly star-forming galaxies at the low redshifts.
We first make a simple assumption that the emission is cooling radiation emitted by cold accretion flows in the form of cold, dense gas.  Here we ignore that some fraction of the Ly$\alpha$ emission is predicted to be produced through resonant scattering of Ly$\alpha$ from inner galactic regions into the halo, and instead assume that all Ly$\alpha$ emission is produced in situ, which may cause us to overestimate the extended H$\alpha$ emission.
In this case, the Ly$\alpha$ emission may be produced primarily from collisional excitation of the gas, rather than recombination.  
Specifically, we i) assume the location where the Ly$\alpha$ and H$\alpha$ emission originates is the same,
ii) assume the ratio of emissivity for H$\alpha$ to Ly$\alpha$ for collisional excitation, iii) correct for cosmological effects on the luminosity,
and iv) ignore resonant scattering of Ly$\alpha$.
Note that the emissivity ratio for H$\alpha$ to Ly$\alpha$ for collisionally excited gas is $\approx$ 1/100 \citep[][]{dijk14}.
Using this method, we estimate that at $\approx$ 80 kpc, the limit out to which \citet{stei11} observe, the surface brightness in H$\alpha$ is $\approx 1.6\times10^{-19}$ ergs s$^{-1}$ arcsec$^{-2}$ cm$^{-2}$ or $\approx 2250$ photons  cm$^{-2}$ sr$^{-1}$ s$^{-1}$.  
This is roughly consistent with the azimuthally averaged radial profiles of the high mass galaxies (m$_{gal}>10^{11} $M$_{\odot}$) from the EAGLE simulation, shown in Fig.~\ref{FigHalphaProfile}.

If the Ly$\alpha$ emission in extended halos is mainly originating from photo-ionized gas, we also need to account for changes in the star-formation rate and lowering of the UV ionizing background (in the case that the Ly$\alpha$ emission originates from UV background-ionized gas).  A simple method of scaling from basic physical processes will not suffice in this case, so instead we turn to the EAGLE simulation.

The azimuthally averaged radial profiles of the CGM of galaxies in EAGLE (see Fig.~\ref{FigHalphaProfile}) allow us to estimate the surface brightness of H$\alpha$ emission from the extended halos.
For each mass bin, the median H$\alpha$ surface brightness is, respectively, SB$_{CGM,inner}\approx$~160, 4800, and 3000 photons  cm$^{-2}$ sr$^{-1}$ s$^{-1}$ at the inner edge of the CGM.  Interestingly, the inner edge of the CGM defined by $\sim$5$\times r_{h, star}$ is brighter for the middle mass bin rather than the largest mass bin.  
While this may be due to small number statistics (there are 19 galaxies  with m$_{gal}>10^{11} $M$_{\odot}$ contained in the 100 cMpc x 100 cMpc x 20 cMpc EAGLE simulation box considered here), this could indicate a difference in build-up mechanism (i.e. gas falls more directly into the smaller mass galaxies and builds up to create a steeper profile, whereas pressure support of virialized gas in the halos of larger mass galaxies, and the heating of infalling gas, prevents gas from falling directly into the galaxy causing a shallower density profile of gas and translating into a shallower slope for the H-alpha surface brightness).  In any case, this demonstrates that it is not necessary to focus on extremely massive galaxies to detect the CGM in H$\alpha$ emission, as galaxies in the middle mass bin are just as bright.

The median surface brightness of the profiles shown in Fig.~\ref{FigHalphaProfile} drops off quickly, falling to $\sim$~1 photon cm$^{-2}$ sr$^{-1}$ s$^{-1}$ by $\approx$~0.2, 0.4, and 0.3 R/R$_{200}$  for the lowest, middle, and highest mass bins, respectively.  As can be seen from the individual galaxy profiles, there are exceptions to the median profile, and indeed, galaxy (iii) in the lowest mass bin is brighter by two orders of magnitude than the median brightness at $\approx$~0.2 R/R$_{200}$.   
Using the largest mass bin at a radius of 100 kpc as the reference point for estimating the H$\alpha$ surface brightness in the extended halo yields a surface brightness from EAGLE of SB$_{100kpc} \approx 700$ photons cm$^{-2}$ sr$^{-1}$ s$^{-1}$.

Similar results to those just presented were found by \citet{vand13}, who used the OWLS cosmological simulations \citep{scha10} to investigate the surface brightness of galactic halos at low redshift.  These authors found SB$_{100kpc} \approx 600$ photons cm$^{-2}$ sr$^{-1}$ s$^{-1}$ for galaxies with halo masses of $10^{13} <$ M$_{halo}$/M$_{\odot}$ $< 10^{14}$.
The two simulations give surface brightness estimates from the H$\alpha$ radial surface brightness profiles that agree well within the scatter.
The H$\alpha$ surface brightness of the hot extended halos of galaxies from the simulations are about four times less luminous than predicted by directly translating from observations of Ly$\alpha$ surface brightness at high redshift. 

Clearly, detecting H$\alpha$ emission out to R$_{200}$ is not feasible for current instruments, but even reaching a fraction of the way into the CGM can provide important constraints on the gas.
For azimuthal averaging at the inner edge of the CGM (corresponding to radii of $\approx$~15 -- 35 kpc for the different mass ranges), the resulting binning corresponds to $\approx$~50 -- 100 arcsec scale features (assuming that the galaxy is at a distance of $\approx$~50 Mpc).  
The signal-to-noise as a function of exposure time for 50 arcsec and 100 arcsec features is shown in the left and middle panels of Fig.~\ref{FigSNR}, respectively, which shows that for these surface brightnesses, H$\alpha$ emission out to the inner edge of the CGM of a galaxy can be detected with 5$\sigma$ confidence in only $\sim$ 1 hour with Dragonfly (for an average galaxy with m$_{gal}>10^{10} $M$_{\odot}$).  

Taking the H$\alpha$ surface brightness estimate from EAGLE, the required exposure time for Dragonfly to measure the emission out to a radius of 100 kpc is $\approx$~40 hours (see the right panel of Fig.~\ref{FigSNR}; azimuthal averaging at radii of 100 kpc corresponds to binning to $\approx$~150 arcsec for galaxies $\approx$~50 Mpc away).
The estimated H$\alpha$ surface brightness from the order-of-magnitude calculation first presented would require an exposure time of $\approx$~3 hours.
Both of these cases are achievable.

\begin{figure*}
\centering
\includegraphics[width=1.0\linewidth]{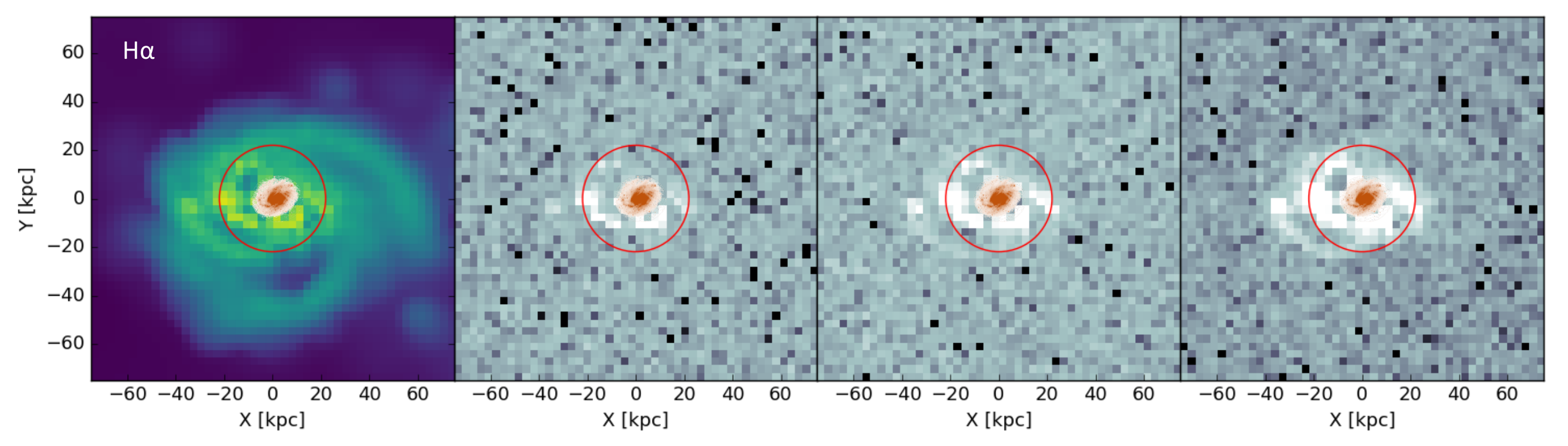}
\caption{ A sample galaxy from the EAGLE simulation (galaxy (iii) from Fig.~\ref{FigFOVwithgalcutouts}) is shown here in the left-most panel.  Superimposed upon the image is an inset of an actual galaxy (NGC 300 from the DSS) to demonstrate the spatial scale of the gaseous structure.  The inset image of NGC 300 has been spatially scaled to match the scale length of the simulation \citep[assuming that the half-stellar mass radius and half-light radius of a galaxy are roughly equal; e.g.][]{szom11}.  The red circle drawn on the image corresponds to 5$\times$ the half-stellar mass radius ($r_{h,star}$) of the EAGLE galaxy.  This radius corresponds to the typical scale we would mask to exclude gas inside the galaxy, and leaves only gas surrounding the galaxy in the CGM.
Mock Dragonfly observations of the sample galaxy are shown in the second through fourth plot.  The second (third; fourth) plot corresponds to an observation with an exposure time of 10 hours (100 hours; 1000 hours) with Dragonfly.  
}
\label{FigMockobsCutout}
\end{figure*}

\subsubsection{Predicted Visibility of Gas Streaming Into and Around Galaxies}

It is clear from the zoom-ins to the CGM of galaxies in Fig.~\ref{FigFOVwithgalcutouts} that circularly averaging the halos of galaxies does not capture the richness of their gas distributions, as non-axi-symmetric inflows and clumpiness exist in the CGM.  
Accretion onto galaxies is predicted to occur through two channels, dubbed ``cold" and ``hot" mode.  The hot mode of accretion is the standard picture, where infalling gas shock heats to near the virial temperature and
then cools radiatively \citep[e.g.][]{rees77,whit91}.  In recent simulations, most accretion is found to actually occur through the cold mode, where dense streams of gas survive infall without being shock-heated, allowing the cool gas to fall in at roughly the free-fall timescale \citep[e.g.][]{kere09,deke09,vand11a}.

To investigate the feasibility of direct imaging of gas flowing around and into galaxies, we take the H$\alpha$ surface brightness projections from the EAGLE simulation directly and add noise to simulate a mock observation.  
Using Dragonfly specifications, we create mock observations by adding sky background and shot noise as well as electronic noise (read out and dark current) to the simulation.  One design feature of Dragonfly is the minimization of scattered light in the optics, resulting in a unique point spread function with a drop of $\sim$ 7 orders of magnitude in flux from the center to a radius of 100 arcsec \citep[][]{abra14}.  The brightest star-forming regions in the EAGLE simulation are $\sim10^8$~photons~cm$^{-2}$~sr$^{-1}$~s$^{-1}$, therefore at an angular distance of 100 arcsec, the scattered light is $\sim10$~photons~cm$^{-2}$~sr$^{-1}$~s$^{-1}$.
The extended wings of the Dragonfly point spread function have not been fully characterized,
 but the point spread function can be reasonably well-approximated by a double Moffat profile with an aureole component \citep[as described in e.g.][]{raci96}.  We convolve the simulation with this point spread function (taking the entire projected Dragonfly field-of-view then cutting out the region of interest) to approximate the scattering of light we would observe.

In Fig.~\ref{FigMockobsCutout}, we show a sample galaxy from the EAGLE simulation in the left panel.  This galaxy has a gaseous halo of H$\alpha$-emitting gas that appears to be spiraling inwards, but could also be a gas disk that extends far out into the CGM.  Note that the spectral resolution of Dragonfly narrow-band filters is not high enough to differentiate between inward-or-outwards motions so emission from all dense streams and clumps of gas (whether infalling or outflowing) will be captured.  The second through fourth panels of this Figure show mock observations of the simulated data in the left-most panel, with different exposure times.  In the second (third; fourth) panel, the exposure time used to create the mock observations is 10 hours (100 hours; 1000 hours) with Dragonfly.  The pixel scale is 3.125 kpc or $\approx$13 arcsec at the projected distance (for reference, Dragonfly's angular resolution is 2.8 arcsec).  In the projected Dragonfly field-of-view, the star-forming regions have surface brightnesses up to $\sim10^{6.5}$~photons~cm$^{-2}$~sr$^{-1}$~s$^{-1}$, resulting in scattered light of $\sim10^{1.5}$~photons~cm$^{-2}$~sr$^{-1}$~s$^{-1}$ at the inner edge of the CGM, which is about an order of magnitude fainter than the brightness of the CGM gas emission.
In each panel, we also include an inset of an actual galaxy, NGC 300 which has been scaled spatially to match the scale length of the simulation, assuming that the half-stellar mass radius and half-light radius are roughly equal\footnote{For reference, the mass and size of the EAGLE galaxy and NGC 300 are not identical:  the EAGLE galaxy has a stellar mass of $\sim5\times10^9M_{\odot}$ and r$_{h,star}\sim4.4$kpc whereas NGC 300 has a stellar mass of $\sim2.1\times10^9M_{\odot}$ (assuming a M/L ratio of 1) and r$_{h,light}\sim3.0$kpc \citep{mcco12}.}.  One can see that the spiraling gas structure extends much farther than the disk of the galaxy:  the red circle corresponds to what is considered to be the edge of a galaxy in stellar light (5$\times r_{h,star}$).
Each panel of Fig.~\ref{FigMockobsCutout} has a red circle at the radius corresponding to 5 $\times$ the half-stellar mass radius ($r_{h,star}$) of the EAGLE galaxy, as for the galaxy cutouts from Fig.~\ref{FigFOVwithgalcutouts}.
Based on Fig.~\ref{FigMockobsCutout}, we conclude that just 10 hours of integration with Dragonfly will allow us to directly observe dense regions of gas outside the outermost limit of the edge of the galaxy (defined by the maximum extent of stellar light) without azimuthally averaging.  In very long (100 to 1000 hour) integrations,
more of the emission is captured, but the emission is so faint that even heroic integrations do not fully reveal the gas.  The outskirts of the gas in the CGM of this mock observation, may, however, be observable with azimuthal averaging (as was described in the previous Section).

\subsubsection{Predicted Visibility of Photoluminescence from the Ultraviolet Background}
In the EAGLE simulations, gas in the CGM and IGM fragments into clouds or clumps, which may be related to so-called ``dark clouds" or ``dark galaxies''.  Recent HI 21 cm surveys have uncovered many ``dark galaxy'' candidates, which are HI clouds with no detected optical counterparts of significant association \citep[for a recent summary, see][]{tayl16} and similar candidates have been found through Ly$\alpha$ emission around high redshift quasars with MUSE \citep{mari18}.
Fluorescent line emission induced by the cosmic ultraviolet background (UVB) from optically thick (to ionizing radiation) HII ``skins" of such intergalactic clouds has never been observed but has long been predicted, though \citet{mari18} observed Ly$\alpha$ fluorescent emission from dark galaxies that is most likely quasar-induced and \citet{fuma17} observed H$\alpha$ fluorescent emission from the disk of a galaxy that is most likely UVB-induced.  
Observations of fluorescent emission from true intergalactic dark galaxies/clouds would place very strong constraints on the (local and/or global) UV ionizing background, which is currently ill-constrained.    Many of these dark HI clouds are $>$100 kpc from their nearest galaxy and, as such, would make good candidates for detecting H$\alpha$ fluorescence.
It is important to know the UV background intensity, since many predictions for CGM absorption and emission depend on it.

Studies of UVB-induced H$\alpha$ emission from dark HI clouds in the local Universe are limited, with one example being that of \citet{dona95}, who undertook narrow-band H$\alpha$ imaging on three intergalactic HI clouds in an attempt to measure the UV ionizing background, probing down to surface brightnesses of $\sim10^{-18}$~erg~s$^{-1}$~cm$^{-2}$~sr$^{-1}$.  
They found non-detections for their two targets that were isolated from any galaxies, and thus more likely to emit H$\alpha$ only through UV background-ionization.
To estimate the required exposure time to observe UVB-induced H$\alpha$ emission, we use estimates of the ionizing UV background and radiative transfer physics to describe the excitation of H clouds and the intensity of the line emission that would result.  Assuming case B recombination at T$\sim10^4$K, about 45\% of the incident ionizing photons result in H$\alpha$ photons \citep{oste06}, and the flux of H$\alpha$ can be estimated to be:  
\begin{equation}
\Phi_{H\alpha} = \frac{J_0 f_{H\alpha}}{h f_g f_s}
\end{equation}
where $J_0$ is the UV ionizing background, $h$ is Planck's constant, $f_g$ is a geometrical correction factor, $f_{H\alpha}\approx0.45$, and $f_s$ is an adjustment for the spectral shape of the ionizing background.  Assuming the cloud is optically thin to H$\alpha$ photons and illuminated by the UVB on all sides, the H$\alpha$ flux depends on the ratio of the clouds surface area to its projected area, which is accounted for by the geometrical factor, f$_g$ \citep{stoc91}.  Since $f_g$ and $f_s$ are unknown, we consider a best-case scenario \citep[i.e. spherical cloud, high ionization background; $J_0~\approx~10^{-22}$ erg s$^{-1}$ cm$^{-2}$ sr$^{-1}$ Hz$^{-1}$ and $f_s~\cdot~f_g~\approx$~1; e.g.][]{fauc09,dona95}, a nominal scenario \citep[i.e. irregularly shaped clouds, high ionization background; $J_0 \approx 2\times10^{-22}$ erg s$^{-1}$ cm$^{-2}$ sr$^{-1}$ Hz$^{-1}$ and $f_s\cdot f_g \approx$ 3.26; e.g.][]{fauc09,dona95}, a pessimistic scenario \citep[i.e. spherical cloud, low ionization background; J$_0 \approx 10^{-23}$ erg s$^{-1}$ cm$^{-2}$ sr$^{-1}$ Hz$^{-1}$ and $f_s\cdot f_g \approx$ 1; e.g.][]{haar12,dona95}, and a worst-case scenario \citep[i.e. irregularly shaped clouds, low ionization background; J$_0 \approx 10^{-23}$ erg s$^{-1}$ cm$^{-2}$ sr$^{-1}$ Hz$^{-1}$ and $f_s\cdot f_g \approx$ 3.26; e.g.][]{haar12,dona95}.

\begin{figure*}
\centering
\includegraphics[width=1.0\linewidth]{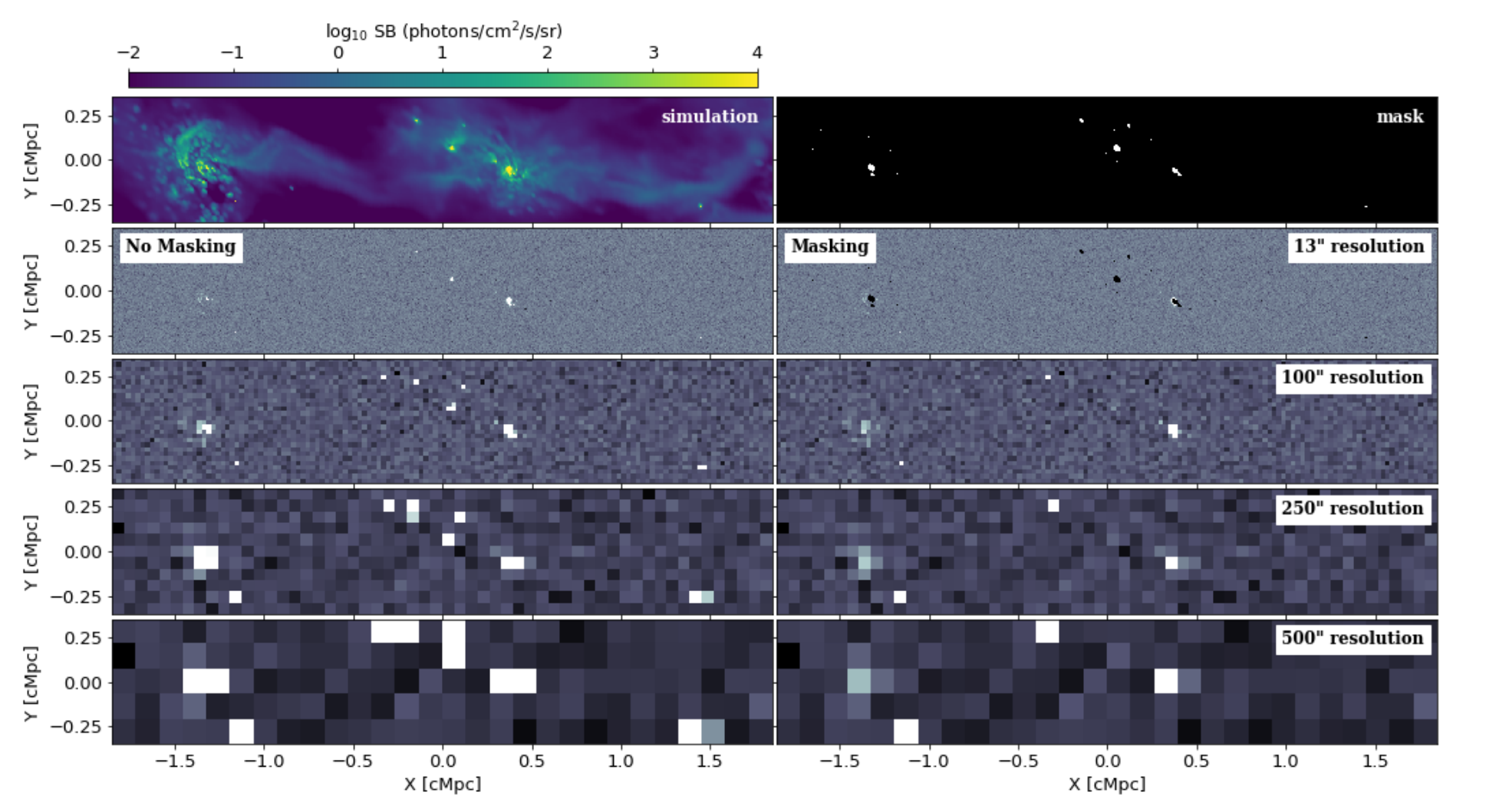}
\caption{ The raw EAGLE data for a filament is shown in the top-left panel.  In the top-right panel, a mask for the filament that was created by masking each galaxy in the field-of-view out to  5 $\times r_{h, stellar mass}$ is shown.
Mock Dragonfly observations of the filament are shown in the second through fifth row of plots, with an exposure time of 1000 hours.  The second row of plots show the highest resolution of the simulation (13" when projected to a distance of 50Mpc from us).  The third (fourth; fifth) row of plots is binned to a resolution of 100" (250"; 500") when projected to a distance of 50 Mpc.
In the left column, no masking of the galaxies in the filament is performed, while on the right, the mask shown in the top-right panel is used to mask out the galaxies in the simulation before binning the data.  Though the filamentary structure itself remains elusive, it is clear that there are bright sources of H$\alpha$ emission outside of galaxies. }
\label{FigBinning}
\end{figure*}

With a signal-to-noise calculated by binning over the number of pixels corresponding to an angular size of 2'x2' \citep[following methods of ][]{dona95}, Dragonfly can reach S/N $\approx$~5 in $\approx$~15 minutes and $\approx$~2.5 hours for the best-case and nominal case scenarios, respectively. The integration time increases to tens of hours for the pessimistic case, and up to a thousand hours for the worst case.  This estimate does not take into account limb-brightening, which boosts the radiation at the edges of the clouds, and may allow one to be slightly more optimistic than the numbers just presented.

Photoluminescence by the UVB can also be targeted by measuring H$\alpha$ emission from the edges of disks of late-type galaxies.
\citet{fuma17} used MUSE to detect H$\alpha$ emission in the outskirts of the galactic disk of UGC 7321 down to surface brightnesses of $\sim1\times10^{-19}\, {\rm erg}\, {\rm cm}^{-2}\,  {\rm s}^{-1}\, {\rm arcsec}^{-2}$ or $\sim$~1400 photons s$^{-1}$ cm$^{-2}$ sr$^{-1}$ and provided constraints on the UVB at z = 0.
Binning to 100 arcsec scale features (which is doable with the large field-of-view of Dragonfly), we predict that this surface brightness can be detected at 5$\sigma$ in $\approx$~15 hours of integration (see the middle panel of Fig.~\ref{FigSNR}) with Dragonfly, allowing similar constraints on the UVB in the local Universe to be made.

\subsection{Predicted Visibility of the Warm-Hot IGM}

While imaging of extended emission from cooling CGM gas in galaxy halos would be extraordinarily interesting, there is no doubt that the `holy grail' would be the detection of gas emission from outside halos and in the IGM itself. 

While monumentally difficult (as we will show), the most spectacular observation would be to directly image the IGM in the cosmic web. 
The simplest analytical arguments suggest that this observation is so difficult as to be effectively hopeless, though we will show that numerical predictions are not quite as pessimistic.

At low redshift, the filamentary IGM is predicted to be mainly collisionally ionized \citep[e.g.][]{bert08}, so emission occurs via radiative cooling, as discussed in \citet{bert13}.  To determine the amount of energy emitted in lines detectable by Dragonfly, we need an estimate of the cosmic web density and mass.
As a first estimate, we imagine that the IGM is simply gas at the mean density in the Universe with an average temperature of T~$\sim10^5$~K (targeting collisionally ionized gas).
The mean density of the Universe corresponds to a hydrogen number density of $\langle n_H\rangle\sim4\times10^{-7}$ cm$^{-3}$ at z $\sim$ 0.
 We take a ballpark estimate of the width of IGM filaments from the simulations of L $\sim$ 0.5 Mpc $\sim 1.5 \times 10^{24}$ cm.  This corresponds to a hydrogen column density $N_H\sim10^{18}$ cm$^{-2}$.  The emission measure (EM) of the IGM filaments can be approximated as EM~=~$\int n_e^2 ds \approx n_e^2 L$, where the integral is evaluated over the length scale of the filament.
The effective recombination rate coefficient for H$\alpha$ emission can be calculated with Equation 14.8 of \citet{drai11}, which yields $\alpha^{eff}_{H\alpha} \approx 1.13\times10^{-14}$ cm$^3$ s$^{-1}$ at temperature T  $\sim10^5$ K.  This rate coefficient is calculated assuming case B recombination, which may not be strictly true for the IGM but suffices for a crude estimate.
The emission rate of H$\alpha$ photons is F$_{H\alpha}~=~\alpha^{eff}_{H\alpha}~EM~\approx~0.006$~ph~cm$^{-2}$~s$^{-1}$.  The surface brightness, $F/\Omega$, is then calculated as SB$_{H\alpha}~=~F~/~(4\pi)~\approx~0.0005$~ph~cm$^{-2}$~s$^{-1}$~sr$^{-1}$.  Clearly, this is extremely faint.

To better approximate filaments in the IGM, we can reasonably assume an average density for the IGM of n$_H\times$10 \citep{bert08,mcqu2015}.  Following the calculation outlined above, we arrive at a surface brightness SB$_{H\alpha} \approx$ 0.5 photons cm$^{-2}$ s$^{-1}$ sr$^{-1}$, which is still very, very faint -- about 1000$\times$ fainter than the extended halos just considered, meaning they would take $\sim$ a million times longer to image!  

A somewhat more optimistic picture emerges if we treat the IGM as a multiphase medium with clumps even denser than n$_H\times$10.  EAGLE allows us to explore this (arguably more realistic) picture of the IGM and its emission.
In the top left panel of Figure~\ref{FigBinning} a zoom in on an example filament of the IGM from EAGLE is displayed.  In the EAGLE simulation, the H$\alpha$ surface brightness of the gas in the IGM ranges from $\sim$100 photon cm$^{-2}$ s$^{-1}$ sr$^{-1}$ (in dense regions near galaxies) to 0.1 photon cm$^{-2}$ s$^{-1}$ sr$^{-1}$ (in isolated, diffuse filaments).  This is consistent with that estimated from the order-of-magnitude approximation for the isolated regions where the gas is extremely diffuse.

In Figure~\ref{FigBinning}, mock Dragonfly observations are plotted for an example filament of the IGM from EAGLE (indicated by the white dashed box in Fig.~\ref{FigMaps}).  The mock observations are created by adding noise and convolving with the Dragonfly point spread function as described in Section 4.1.2. 
The top-left panel of Figure~\ref{FigBinning} shows the raw EAGLE data for the example filament.  
Before we create the mock observations, we make a mask for the filament to mask out emission from the galaxies.  In the top-right panel, the mask is shown:  each galaxy in the filament is masked out to a radius of 5 $\times r_{h, star}$ for that galaxy.
Mock Dragonfly observations of the filament are shown in the second through fifth row of plots, each with an exposure time of 1000 hours.  The second row of plots shows the highest resolution of the simulation ($\sim$13" when projected to a distance of 50Mpc from us).  The third (fourth; fifth) row of plots is binned to a resolution of $\sim$100" (250"; 500").
In the left column, we bin the data without using the mask.
We compare this with the right column, where the mask shown in the top-right panel was applied before binning the data, thus in the right column, the emission peaks in the mock observations are nominally from gas outside of galaxies.
We confirm this supposition in Appendix A2.
Though the filamentary structure itself remains elusive in this mock observation, it is clear that there are bright sources of H$\alpha$ emission outside of galaxies. 

It should be noted that in the EAGLE simulation, portions of the filamentary IGM emission reach surface brightnesses $\sim1$~photon~cm$^{-2}$~sr$^{-1}$~s$^{-1}$, which is of order the brightness of scattered light emission from star-forming regions (as approximated by the characterization of the Dragonfly point spread function; see discussion in Section 4.1.2).
To attain the goal of imaging IGM filaments, down to the surface brightness of 1 photon cm$^{-2}$ s$^{-1}$ sr$^{-1}$ that the EAGLE simulations suggest, extreme binning and upgrades to Dragonfly (e.g. more lenses, new cameras) are necessary.  
As is shown in Fig.~\ref{FigSNR}, even with azimuthal averaging (or extreme binning), a surface brightness of 100  photons cm$^{-2}$ s$^{-1}$ sr$^{-1}$, is barely reachable in 1000 hours of exposure time with Dragonfly as it stands.  

\section{Discussion \& Conclusions}

The hydrodynamical simulations presented here show that direct imaging (at visible wavelengths) of cooling emission from the local circumgalactic medium is now a practical possibility. Such observations are made practical by technical advances in spatially resolved spectroscopy \citep{matu10,baco10,morr12,quir14} and low-surface brightness imaging \citep[][Lokhorst et al.\ in preparation]{abra14}. With appropriate control of systematics (see below), it will be possible to extend the most recent generation of local H$\alpha$ galaxy surveys \citep[e.g.][]{meur06, kenn08, gava12, bose15,vans16}, as well as deeper studies of star formation in local galaxies \citep[e.g.][]{lee16}, out to radii where line emission becomes dominated by gas cooling rather than by photoionization. Such observations would provide a powerful extension of existing techniques for the exploration of the disk-halo interface of galaxies (and of the CGM generally), which have relied mainly on absorption-line spectroscopy using pencil-beam surveys. Since direct imaging at visible wavelengths probes gaseous material at temperatures and densities typical of baryons in the local Universe, these observations would usefully augment investigations focusing on gas in other phases probed by radio (cold gas) and X-ray (hot gas) wavelengths.

The easiest way to characterize warm-hot gaseous emission in the local CGM will be to target the extended halos of galaxies, since the signal-to-noise level of these structures can be boosted by azimuthal averaging. In only one hour of integration time, a mosaic telescope similar to the Dragonfly Telephoto Array with a set of 3 nm bandpass narrow-band filters\footnote{An experimental setup with 3nm filters is under construction and preliminary results will be presented in a companion paper. This imager is based on a six-lens telephoto array with full aperture filters. Central wavelengths are chosen to avoid galactic H$\alpha$ emission, and a differential background subtraction technique (based on tilting the filters to shift their bandpasses) is used to minimize sky contamination. The interested reader is referred to Lokhorst et al.\ in preparation for details.} could readily observe H$\alpha$ emission at the inner edge of the CGM (for which we have adopted the definition of 5$\times r_{h,\star}$), where the median H$\alpha$ emission is approximately 3000 -- 5000 photons s$^{-1}$ cm$^{-2}$ sr$^{-1}$ for galaxies at redshift $z$ $\sim$ 0 with stellar masses $\geq$10$^{10}$ M$_{\odot}$ (see Section 3.1.1). This radius corresponds to the outermost distance at which starlight has been detected in the disks of galaxies. Pushing to more ambitious integration times would allow one to probe out to radii well beyond those at which stars are seen in local disk galaxies. For example, EAGLE predicts that a narrow-band imaging telescope with similar characteristics to Dragonfly would be able to map radial profiles down to surface brightnesses of $\sim$700 photons s$^{-1}$ cm$^{-2}$ sr$^{-1}$ with exposure times of around 40 hours. This would allow the detection of H$\alpha$ emission out to radii of around 100 kpc (for a galaxy with stellar mass $\geq$10$^{11}$ M$_{\odot}$). 

These predictions are based on EAGLE, but one can obtain similar numbers using empirical arguments. For example, we have shown that the Ly$\alpha$ halo surface brightness profile measured by \citet{stei11} at high-redshift can be used to predict the corresponding H$\alpha$ surface brightness of the halos of local galaxies.  Assuming the emission is produced through cooling radiation, H$\alpha$ emission would be $\sim$ 4 times stronger than that predicted by the EAGLE simulation. This may simply be a reflection of the fact that the observed emission is the product of both cooling radiation and photo-ionization by star-forming regions or even by the extragalactic UV-background\footnote{Measurement of Ly$\alpha$ at low surface brightness is complicated by the fact that Ly$\alpha$ is a resonantly scattering line, so it is possible for star-forming regions to light up emission in the outskirts of the galaxy, decoupling the line strength from the gas density \citep[e.g.][]{fauc10}. H$\alpha$ is not a resonant scatterer, so it will trace the gas density more closely.}. In any case, the main point is that a local star-forming galaxy with properties similar to those of the (admittedly fairly extreme) high-redshift objects studied by \citet{stei11} would almost certainly show an H$\alpha$ halo that would be readily detectable by a narrow-band imager optimized for the detection of low-surface brightness structures. 

Moving beyond the investigation of axisymmetric structures makes the prospects for observing emission from the local CGM/IGM more nuanced. It would be extremely challenging to detect gaseous emission from the largest scale filamentary structure in the local Universe (i.e. from material very distant from halos and confined only by the gravity of the cosmic web). The surface brightness of H$\alpha$ emission from this filamentary emission is extremely low, at only a few photons s$^{-1}$ cm$^{-2}$ sr$^{-1}$.  Even when using very narrow bandpass filters (such as the 3 nm bandwidth filters described in Lokhorst et al. in prep.) and binning to extremely low spatial resolution ($\sim$ 100 arcsec FWHM beams), a Dragonfly-like telescope would require integration times of tens of thousands of hours to trace out directly the structure of the cosmic web over something like ten degrees of sky. This seems hopeless, but at present the world's largest mosaic telescope (Dragonfly) has the effective aperture of only a 1m telescope. The effective aperture of small telescope arrays can be scaled up relatively easily, and because they build up aperture by averaging over many beams, control over systematics grows in lock-step with the size of the array. There is some hope that in the future the direct detection of even the `deep' cosmic web will fall within the reach of a large mosaic telescope array. In the meantime, statistical methods may be used to augment direct imaging approaches for investigating the cosmic web, e.g. via cross-correlation of extended H$\alpha$ or [OIII] emission with the positions of galaxies, as was done by \citet{crof16} with Ly$\alpha$ emitters and quasars in the Sloan Digital Sky Survey at intermediate redshift. 

Focusing on volumes of space closer to galaxies brings us to a very interesting (and observationally tractable) regime, where the CGM of the galaxies is dominated by non-axisymmetric inflowing gas \citep[e.g.][]{mart14, mart14b}. The present paper suggests that detecting the diffuse gas in streams is now a realistic prospect. The requisite observations would take a Dragonfly-like telescope significant (but realistically achievable) amounts of time --- EAGLE suggests that un-binned integration times range from tens of hours with optimistic assumptions to thousands of hours with pessimistic assumptions. Observations of the more diffuse components of the CGM could perhaps be undertaken with extreme binning, but a better strategy might be to focus on the detection of dense pockets in these streams. Clumps of gas in streams may be related to dark HI clouds which have been observed near galaxies and have no stellar counterparts. As suggested by \cite{dona95}, if these clouds are far enough away from galaxies, they will be solely illuminated by the UV-ionizing background and observations of line emission from them would allow an estimate to be made of the local UV-ionizing background. The UV-ionizing background at redshift $z$ $\sim$ 0 is currently only constrained to within two orders of magnitude \citep[e.g. see][for a recent summary]{fuma17} and placing better constraints on this important parameter would appear to be both relatively straightforward and of great interest.

The central message of the present paper is this: {\em direct imaging at visible wavelengths of gas inflow and feedback at the disk-halo interface of local galaxies is now a tractable observational problem}. Parallel progress mapping Ly$\alpha$ emission from gas in the CGM/IGM is being made at high-redshifts using spatially resolved spectrometers such as MUSE and KCWI \citep{mart14b,mart16,wiso16}, with especial success targeting quasar-illuminated gas \citep{mart14,bori16}.  Additionally, in even more local environments than those considered here, progress has been make in the detections of baryons in the halo of the Milky Way through stacking of Sloan Digital Sky Survey spectra \citep{zhan17}.  The significance of this progress is that we will soon be able to undertake observations which  characterize directly the relationship between galaxy growth (both in gas and in stars) and feedback (inflows and outflows). In doing so, we will also be imaging the hidden dominant component of the Universe's baryons that, together with dark matter, acts behind the scenes to control the formation of galaxies.

\acknowledgments
We thank C. Matzner for useful conversations about the analytical calculations in this paper and the anonymous referee whose helpful comments lead to improvements in the manuscript.
We are thankful for contributions from the Dunlap Institute (funded through an endowment established by the David Dunlap family and the University of Toronto).

\bibliographystyle{apj}

\begin{thebibliography}{}
\expandafter\ifx\csname natexlab\endcsname\relax\def\natexlab#1{#1}\fi

\bibitem[{{Abraham} \& {van Dokkum}(2014)}]{abra14}
{Abraham}, R.~G., \& {van Dokkum}, P.~G. 2014, \pasp, 126, 55

\bibitem[{{Allende Prieto} {et~al.}(2001){Allende Prieto}, {Lambert}, \&
  {Asplund}}]{alle01}
{Allende Prieto}, C., {Lambert}, D.~L., \& {Asplund}, M. 2001, \apjl, 556, L63

\bibitem[{{Allende Prieto} {et~al.}(2002){Allende Prieto}, {Lambert}, \&
  {Asplund}}]{alle02}
---. 2002, \apjl, 573, L137

\bibitem[{{Altay} {et~al.}(2011){Altay}, {Theuns}, {Schaye}, {Crighton}, \&
  {Dalla Vecchia}}]{alta11}
{Altay}, G., {Theuns}, T., {Schaye}, J., {Crighton}, N.~H.~M., \& {Dalla
  Vecchia}, C. 2011, \apjl, 737, L37

\bibitem[{{Arrigoni Battaia} {et~al.}(2018){Arrigoni Battaia}, {Prochaska},
  {Hennawi}, {Obreja}, {Buck}, {Cantalupo}, {Dutton}, \& {Macci{\`o}}}]{arri18}
{Arrigoni Battaia}, F., {Prochaska}, J.~X., {Hennawi}, J.~F., {et~al.} 2018,
  \mnras, 473, 3907

\bibitem[{{Bacon} {et~al.}(2010){Bacon}, {Accardo}, {Adjali}, {Anwand},
  {Bauer}, {Biswas}, {Blaizot}, {Boudon}, {Brau-Nogue}, {Brinchmann},
  {Caillier}, {Capoani}, {Carollo}, {Contini}, {Couderc}, {Daguis{\'e}},
  {Deiries}, {Delabre}, {Dreizler}, {Dubois}, {Dupieux}, {Dupuy}, {Emsellem},
  {Fechner}, {Fleischmann}, {Fran{\c c}ois}, {Gallou}, {Gharsa}, {Glindemann},
  {Gojak}, {Guiderdoni}, {Hansali}, {Hahn}, {Jarno}, {Kelz}, {Koehler},
  {Kosmalski}, {Laurent}, {Le Floch}, {Lilly}, {Lizon}, {Loupias}, {Manescau},
  {Monstein}, {Nicklas}, {Olaya}, {Pares}, {Pasquini}, {P{\'e}contal-Rousset},
  {Pell{\'o}}, {Petit}, {Popow}, {Reiss}, {Remillieux}, {Renault}, {Roth},
  {Rupprecht}, {Serre}, {Schaye}, {Soucail}, {Steinmetz}, {Streicher}, {Stuik},
  {Valentin}, {Vernet}, {Weilbacher}, {Wisotzki}, \& {Yerle}}]{baco10}
{Bacon}, R., {Accardo}, M., {Adjali}, L., {et~al.} 2010, in \procspie, Vol.
  7735, Ground-based and Airborne Instrumentation for Astronomy III, 773508

\bibitem[{{Bauermeister} {et~al.}(2010){Bauermeister}, {Blitz}, \&
  {Ma}}]{baue10}
{Bauermeister}, A., {Blitz}, L., \& {Ma}, C.-P. 2010, \apj, 717, 323

\bibitem[{{Benn} \& {Ellison}(1998)}]{benn98}
{Benn}, C.~R., \& {Ellison}, S.~L. 1998, \nar, 42, 503

\bibitem[{{Bertone} {et~al.}(2013){Bertone}, {Aguirre}, \& {Schaye}}]{bert13}
{Bertone}, S., {Aguirre}, A., \& {Schaye}, J. 2013, \mnras, 430, 3292

\bibitem[{{Bertone} \& {Schaye}(2012)}]{bert12}
{Bertone}, S., \& {Schaye}, J. 2012, \mnras, 419, 780

\bibitem[{{Bertone} {et~al.}(2010{\natexlab{a}}){Bertone}, {Schaye}, {Booth},
  {Dalla Vecchia}, {Theuns}, \& {Wiersma}}]{bert10b}
{Bertone}, S., {Schaye}, J., {Booth}, C.~M., {et~al.} 2010{\natexlab{a}},
  \mnras, 408, 1120

\bibitem[{{Bertone} {et~al.}(2010{\natexlab{b}}){Bertone}, {Schaye}, {Dalla
  Vecchia}, {Booth}, {Theuns}, \& {Wiersma}}]{bert10a}
{Bertone}, S., {Schaye}, J., {Dalla Vecchia}, C., {et~al.} 2010{\natexlab{b}},
  \mnras, 407, 544

\bibitem[{{Bertone} {et~al.}(2008){Bertone}, {Schaye}, \& {Dolag}}]{bert08}
{Bertone}, S., {Schaye}, J., \& {Dolag}, K. 2008, \ssr, 134, 295

\bibitem[{{Bland-Hawthorn} {et~al.}(2005){Bland-Hawthorn}, {Vlaji{\'c}},
  {Freeman}, \& {Draine}}]{blan05}
{Bland-Hawthorn}, J., {Vlaji{\'c}}, M., {Freeman}, K.~C., \& {Draine}, B.~T.
  2005, \apj, 629, 239

\bibitem[{{Borisova} {et~al.}(2016){Borisova}, {Cantalupo}, {Lilly}, {Marino},
  {Gallego}, {Bacon}, {Blaizot}, {Bouch{\'e}}, {Brinchmann}, {Carollo},
  {Caruana}, {Finley}, {Herenz}, {Richard}, {Schaye}, {Straka}, {Turner},
  {Urrutia}, {Verhamme}, \& {Wisotzki}}]{bori16}
{Borisova}, E., {Cantalupo}, S., {Lilly}, S.~J., {et~al.} 2016, \apj, 831, 39

\bibitem[{{Boselli} {et~al.}(2015){Boselli}, {Fossati}, {Gavazzi}, {Ciesla},
  {Buat}, {Boissier}, \& {Hughes}}]{bose15}
{Boselli}, A., {Fossati}, M., {Gavazzi}, G., {et~al.} 2015, \aap, 579, A102

\bibitem[{{Cai} {et~al.}(2018){Cai}, {Hamden}, {Matuszewski}, {Prochaska},
  {Li}, {Cantalupo}, {Arrigoni Battaia}, {Martin}, {Neill}, {O'Sullivan},
  {Wang}, {Moore}, \& {Morrissey}}]{cai18}
{Cai}, Z., {Hamden}, E., {Matuszewski}, M., {et~al.} 2018, \apjl, 861, L3

\bibitem[{{Crain} {et~al.}(2015){Crain}, {Schaye}, {Bower}, {Furlong},
  {Schaller}, {Theuns}, {Dalla Vecchia}, {Frenk}, {McCarthy}, {Helly},
  {Jenkins}, {Rosas-Guevara}, {White}, \& {Trayford}}]{crai15}
{Crain}, R.~A., {Schaye}, J., {Bower}, R.~G., {et~al.} 2015, \mnras, 450, 1937

\bibitem[{{Croft} {et~al.}(2016){Croft}, {Miralda-Escud{\'e}}, {Zheng},
  {Bolton}, {Dawson}, {Peterson}, {York}, {Eisenstein}, {Brinkmann},
  {Brownstein}, {Cen}, {Delubac}, {Font-Ribera}, {Hamilton}, {Lee}, {Myers},
  {Palanque-Delabrouille}, {P{\^a}ris}, {Petitjean}, {Pieri}, {Ross}, {Rossi},
  {Schlegel}, {Schneider}, {Slosar}, {Vazquez}, {Viel}, {Weinberg}, \&
  {Y{\`e}che}}]{crof16}
{Croft}, R.~A.~C., {Miralda-Escud{\'e}}, J., {Zheng}, Z., {et~al.} 2016,
  \mnras, 457, 3541

\bibitem[{{Dalla Vecchia} \& {Schaye}(2012)}]{dall12}
{Dalla Vecchia}, C., \& {Schaye}, J. 2012, \mnras, 426, 140

\bibitem[{{Danforth} {et~al.}(2016){Danforth}, {Keeney}, {Tilton}, {Shull},
  {Stocke}, {Stevans}, {Pieri}, {Savage}, {France}, {Syphers}, {Smith},
  {Green}, {Froning}, {Penton}, \& {Osterman}}]{danf16}
{Danforth}, C.~W., {Keeney}, B.~A., {Tilton}, E.~M., {et~al.} 2016, \apj, 817,
  111

\bibitem[{{Dekel} {et~al.}(2009){Dekel}, {Birnboim}, {Engel}, {Freundlich},
  {Goerdt}, {Mumcuoglu}, {Neistein}, {Pichon}, {Teyssier}, \&
  {Zinger}}]{deke09}
{Dekel}, A., {Birnboim}, Y., {Engel}, G., {et~al.} 2009, \nat, 457, 451

\bibitem[{{Dijkstra}(2014)}]{dijk14}
{Dijkstra}, M. 2014, \pasa, 31, e040

\bibitem[{{Donahue} {et~al.}(1995){Donahue}, {Aldering}, \& {Stocke}}]{dona95}
{Donahue}, M., {Aldering}, G., \& {Stocke}, J.~T. 1995, \apjl, 450, L45

\bibitem[{{Draine}(2011)}]{drai11}
{Draine}, B.~T. 2011, {Physics of the Interstellar and Intergalactic Medium}

\bibitem[{{Emonts} {et~al.}(2018){Emonts}, {Lehnert}, {Dannerbauer}, {De
  Breuck}, {Villar-Mart{\'{\i}}n}, {Miley}, {Allison}, {Gullberg}, {Hatch},
  {Guillard}, {Mao}, \& {Norris}}]{emon18}
{Emonts}, B.~H.~C., {Lehnert}, M.~D., {Dannerbauer}, H., {et~al.} 2018, \mnras,
  477, L60

\bibitem[{{Faucher-Gigu{\`e}re} {et~al.}(2010){Faucher-Gigu{\`e}re}, {Kere{\v
  s}}, {Dijkstra}, {Hernquist}, \& {Zaldarriaga}}]{fauc10}
{Faucher-Gigu{\`e}re}, C.-A., {Kere{\v s}}, D., {Dijkstra}, M., {Hernquist},
  L., \& {Zaldarriaga}, M. 2010, \apj, 725, 633

\bibitem[{{Faucher-Gigu{\`e}re} {et~al.}(2009){Faucher-Gigu{\`e}re}, {Lidz},
  {Zaldarriaga}, \& {Hernquist}}]{fauc09}
{Faucher-Gigu{\`e}re}, C.-A., {Lidz}, A., {Zaldarriaga}, M., \& {Hernquist}, L.
  2009, \apj, 703, 1416

\bibitem[{{Ferland} {et~al.}(1998){Ferland}, {Korista}, {Verner}, {Ferguson},
  {Kingdon}, \& {Verner}}]{ferl98}
{Ferland}, G.~J., {Korista}, K.~T., {Verner}, D.~A., {et~al.} 1998, \pasp, 110,
  761

\bibitem[{{Fumagalli} {et~al.}(2017){Fumagalli}, {Haardt}, {Theuns}, {Morris},
  {Cantalupo}, {Madau}, \& {Fossati}}]{fuma17}
{Fumagalli}, M., {Haardt}, F., {Theuns}, T., {et~al.} 2017, \mnras, 467, 4802

\bibitem[{{Gavazzi} {et~al.}(2012){Gavazzi}, {Fumagalli}, {Galardo},
  {Grossetti}, {Boselli}, {Giovanelli}, {Haynes}, \& {Fabello}}]{gava12}
{Gavazzi}, G., {Fumagalli}, M., {Galardo}, V., {et~al.} 2012, \aap, 545, A16

\bibitem[{{Gnat} \& {Sternberg}(2009)}]{gnat09}
{Gnat}, O., \& {Sternberg}, A. 2009, \apj, 693, 1514

\bibitem[{{Gould} \& {Weinberg}(1996)}]{goul96}
{Gould}, A., \& {Weinberg}, D.~H. 1996, \apj, 468, 462

\bibitem[{{Haardt} \& {Madau}(2001)}]{haar01}
{Haardt}, F., \& {Madau}, P. 2001, in Clusters of Galaxies and the High
  Redshift Universe Observed in X-rays, ed. D.~M. {Neumann} \& J.~T.~V. {Tran},
  64

\bibitem[{{Haardt} \& {Madau}(2012)}]{haar12}
{Haardt}, F., \& {Madau}, P. 2012, \apj, 746, 125

\bibitem[{{Hanuschik}(2003)}]{hanu03}
{Hanuschik}, R.~W. 2003, \aap, 407, 1157

\bibitem[{{Heald} {et~al.}(2011){Heald}, {J{\'o}zsa}, {Serra}, {Zschaechner},
  {Rand}, {Fraternali}, {Oosterloo}, {Walterbos}, {J{\"u}tte}, \&
  {Gentile}}]{heal11}
{Heald}, G., {J{\'o}zsa}, G., {Serra}, P., {et~al.} 2011, \aap, 526, A118

\bibitem[{{Holweger}(2001)}]{holw01}
{Holweger}, H. 2001, in American Institute of Physics Conference Series, Vol.
  598, Joint SOHO/ACE workshop ``Solar and Galactic Composition'', ed. R.~F.
  {Wimmer-Schweingruber}, 23--30

\bibitem[{{Keeney} {et~al.}(2017){Keeney}, {Stocke}, {Danforth}, {Shull},
  {Pratt}, {Froning}, {Green}, {Penton}, \& {Savage}}]{keen17}
{Keeney}, B.~A., {Stocke}, J.~T., {Danforth}, C.~W., {et~al.} 2017, \apjs, 230,
  6

\bibitem[{{Kennicutt} \& {Evans}(2012)}]{kenn12}
{Kennicutt}, R.~C., \& {Evans}, N.~J. 2012, \araa, 50, 531

\bibitem[{{Kennicutt} {et~al.}(2008){Kennicutt}, {Lee}, {Funes}, {J.}, {Sakai},
  \& {Akiyama}}]{kenn08}
{Kennicutt}, Jr., R.~C., {Lee}, J.~C., {Funes}, J.~G., {et~al.} 2008, \apjs,
  178, 247

\bibitem[{{Kere{\v s}} {et~al.}(2009){Kere{\v s}}, {Katz}, {Fardal},
  {Dav{\'e}}, \& {Weinberg}}]{kere09}
{Kere{\v s}}, D., {Katz}, N., {Fardal}, M., {Dav{\'e}}, R., \& {Weinberg},
  D.~H. 2009, \mnras, 395, 160

\bibitem[{{Krisciunas} \& {Schaefer}(1991)}]{kris91}
{Krisciunas}, K., \& {Schaefer}, B.~E. 1991, \pasp, 103, 1033

\bibitem[{{Lee} {et~al.}(2016){Lee}, {Veilleux}, {McDonald}, \&
  {Hilbert}}]{lee16}
{Lee}, J.~C., {Veilleux}, S., {McDonald}, M., \& {Hilbert}, B. 2016, \apj, 817,
  177

\bibitem[{{Leibler} {et~al.}(2018){Leibler}, {Cantalupo}, {Holden}, \&
  {Madau}}]{leib18}
{Leibler}, C.~N., {Cantalupo}, S., {Holden}, B.~P., \& {Madau}, P. 2018,
  \mnras, 480, 2094

\bibitem[{{Lodders}(2003)}]{lodd03}
{Lodders}, K. 2003, \apj, 591, 1220

\bibitem[{{Marino} {et~al.}(2018){Marino}, {Cantalupo}, {Lilly}, {Gallego},
  {Straka}, {Borisova}, {Pezzulli}, {Bacon}, {Brinchmann}, {Carollo},
  {Caruana}, {Conseil}, {Contini}, {Diener}, {Finley}, {Inami}, {Leclercq},
  {Muzahid}, {Richard}, {Schaye}, {Wendt}, \& {Wisotzki}}]{mari18}
{Marino}, R.~A., {Cantalupo}, S., {Lilly}, S.~J., {et~al.} 2018, \apj, 859, 53

\bibitem[{{Martin} {et~al.}(2010){Martin}, {Moore}, {Morrissey}, {Matuszewski},
  {Rahman}, {Adkins}, \& {Epps}}]{mart10}
{Martin}, C., {Moore}, A., {Morrissey}, P., {et~al.} 2010, in \procspie, Vol.
  7735, Ground-based and Airborne Instrumentation for Astronomy III, 77350M

\bibitem[{{Martin} {et~al.}(2013){Martin}, {Shapley}, {Coil}, {Kornei},
  {Murray}, \& {Pancoast}}]{mart13}
{Martin}, C.~L., {Shapley}, A.~E., {Coil}, A.~L., {et~al.} 2013, \apj, 770, 41

\bibitem[{{Martin} {et~al.}(2014{\natexlab{a}}){Martin}, {Chang},
  {Matuszewski}, {Morrissey}, {Rahman}, {Moore}, \& {Steidel}}]{mart14}
{Martin}, D.~C., {Chang}, D., {Matuszewski}, M., {et~al.} 2014{\natexlab{a}},
  \apj, 786, 106

\bibitem[{{Martin} {et~al.}(2014{\natexlab{b}}){Martin}, {Chang},
  {Matuszewski}, {Morrissey}, {Rahman}, {Moore}, {Steidel}, \&
  {Matsuda}}]{mart14b}
---. 2014{\natexlab{b}}, \apj, 786, 107

\bibitem[{{Martin} {et~al.}(2016){Martin}, {Matuszewski}, {Morrissey}, {Neill},
  {Moore}, {Steidel}, \& {Trainor}}]{mart16}
{Martin}, D.~C., {Matuszewski}, M., {Morrissey}, P., {et~al.} 2016, \apjl, 824,
  L5

\bibitem[{{Matsuda} {et~al.}(2012){Matsuda}, {Yamada}, {Hayashino}, {Yamauchi},
  {Nakamura}, {Morimoto}, {Ouchi}, {Ono}, {Umemura}, \& {Mori}}]{mats12}
{Matsuda}, Y., {Yamada}, T., {Hayashino}, T., {et~al.} 2012, \mnras, 425, 878

\bibitem[{{Matuszewski} {et~al.}(2010){Matuszewski}, {Chang}, {Crabill},
  {Martin}, {Moore}, {Morrissey}, \& {Rahman}}]{matu10}
{Matuszewski}, M., {Chang}, D., {Crabill}, R.~M., {et~al.} 2010, in \procspie,
  Vol. 7735, Ground-based and Airborne Instrumentation for Astronomy III,
  77350P

\bibitem[{{McAlpine} {et~al.}(2016){McAlpine}, {Helly}, {Schaller}, {Trayford},
  {Qu}, {Furlong}, {Bower}, {Crain}, {Schaye}, {Theuns}, {Dalla Vecchia},
  {Frenk}, {McCarthy}, {Jenkins}, {Rosas-Guevara}, {White}, {Baes}, {Camps}, \&
  {Lemson}}]{mcal16}
{McAlpine}, S., {Helly}, J.~C., {Schaller}, M., {et~al.} 2016, Astronomy and
  Computing, 15, 72

\bibitem[{{McConnachie}(2012)}]{mcco12}
{McConnachie}, A.~W. 2012, \aj, 144, 4

\bibitem[{{McQuinn}(2016)}]{mcqu2015}
{McQuinn}, M. 2016, \araa, 54, 313

\bibitem[{{Meurer} {et~al.}(2006){Meurer}, {Hanish}, {Ferguson}, {Knezek},
  {Kilborn}, {Putman}, {Smith}, {Koribalski}, {Meyer}, {Oey}, {Ryan-Weber},
  {Zwaan}, {Heckman}, {Kennicutt}, {Lee}, {Webster}, {Bland-Hawthorn},
  {Dopita}, {Freeman}, {Doyle}, {Drinkwater}, {Staveley-Smith}, \&
  {Werk}}]{meur06}
{Meurer}, G.~R., {Hanish}, D.~J., {Ferguson}, H.~C., {et~al.} 2006, \apjs, 165,
  307

\bibitem[{{Momose} {et~al.}(2014){Momose}, {Ouchi}, {Nakajima}, {Ono},
  {Shibuya}, {Shimasaku}, {Yuma}, {Mori}, \& {Umemura}}]{momo14}
{Momose}, R., {Ouchi}, M., {Nakajima}, K., {et~al.} 2014, \mnras, 442, 110

\bibitem[{{Momose} {et~al.}(2016){Momose}, {Ouchi}, {Nakajima}, {Ono},
  {Shibuya}, {Shimasaku}, {Yuma}, {Mori}, \& {Umemura}}]{momo16}
---. 2016, \mnras, 457, 2318

\bibitem[{{Morrissey} {et~al.}(2012){Morrissey}, {Matuszewski}, {Martin},
  {Moore}, {Adkins}, {Epps}, {Bartos}, {Cabak}, {Cowley}, {Davis}, {Delacroix},
  {Fucik}, {Hilliard}, {James}, {Kaye}, {Lingner}, {Neill}, {Pistor},
  {Phillips}, {Rockosi}, \& {Weber}}]{morr12}
{Morrissey}, P., {Matuszewski}, M., {Martin}, C., {et~al.} 2012, in \procspie,
  Vol. 8446, Ground-based and Airborne Instrumentation for Astronomy IV, 844613

\bibitem[{{Moss} {et~al.}(2017){Moss}, {Lockman}, \&
  {McClure-Griffiths}}]{moss17}
{Moss}, V.~A., {Lockman}, F.~J., \& {McClure-Griffiths}, N.~M. 2017, \apj, 834,
  155

\bibitem[{{Nicastro} {et~al.}(2018){Nicastro}, {Kaastra}, {Krongold},
  {Borgani}, {Branchini}, {Cen}, {Dadina}, {Danforth}, {Elvis}, {Fiore},
  {Gupta}, {Mathur}, {Mayya}, {Paerels}, {Piro}, {Rosa-Gonzalez}, {Schaye},
  {Shull}, {Torres-Zafra}, {Wijers}, \& {Zappacosta}}]{nica18}
{Nicastro}, F., {Kaastra}, J., {Krongold}, Y., {et~al.} 2018, \nat, 558, 406

\bibitem[{{Oosterloo} {et~al.}(2007){Oosterloo}, {Fraternali}, \&
  {Sancisi}}]{oost07}
{Oosterloo}, T., {Fraternali}, F., \& {Sancisi}, R. 2007, \aj, 134, 1019

\bibitem[{{Oppenheimer} \& {Schaye}(2013)}]{oppe13}
{Oppenheimer}, B.~D., \& {Schaye}, J. 2013, \mnras, 434, 1043

\bibitem[{{Osterbrock} \& {Ferland}(2006)}]{oste06}
{Osterbrock}, D.~E., \& {Ferland}, G.~J. 2006, {Astrophysics of gaseous nebulae
  and active galactic nuclei}

\bibitem[{{Pingel} {et~al.}(2018){Pingel}, {Pisano}, {Heald}, {Jarrett}, {de
  Blok}, {J{\'o}zsa}, {J{\"u}tte}, {Rand}, {Oosterloo}, \& {Winkel}}]{ping18}
{Pingel}, N.~M., {Pisano}, D.~J., {Heald}, G., {et~al.} 2018, \apj, 865, 36

\bibitem[{{Planck Collaboration} {et~al.}(2014){Planck Collaboration}, {Ade},
  {Aghanim}, {Alves}, {Armitage-Caplan}, {Arnaud}, {Ashdown},
  {Atrio-Barandela}, {Aumont}, {Aussel}, \& et~al.}]{plan14}
{Planck Collaboration}, {Ade}, P.~A.~R., {Aghanim}, N., {et~al.} 2014, \aap,
  571, A1

\bibitem[{{Prochaska} {et~al.}(2017){Prochaska}, {Werk}, {Worseck}, {Tripp},
  {Tumlinson}, {Burchett}, {Fox}, {Fumagalli}, {Lehner}, {Peeples}, \&
  {Tejos}}]{proc17}
{Prochaska}, J.~X., {Werk}, J.~K., {Worseck}, G., {et~al.} 2017, \apj, 837, 169

\bibitem[{{Quiret} {et~al.}(2014){Quiret}, {Milliard}, {Grange}, {Lemaitre},
  {Caillat}, {Belhadi}, \& {Cotel}}]{quir14}
{Quiret}, S., {Milliard}, B., {Grange}, R., {et~al.} 2014, in \procspie, Vol.
  9144, Space Telescopes and Instrumentation 2014: Ultraviolet to Gamma Ray,
  914432

\bibitem[{{Racine}(1996)}]{raci96}
{Racine}, R. 1996, Publications of the Astronomical Society of the Pacific,
  108, 699

\bibitem[{{Rahmati} {et~al.}(2013{\natexlab{a}}){Rahmati}, {Pawlik}, {Rai{\v
  c}evi{\'c}}, \& {Schaye}}]{rahm13}
{Rahmati}, A., {Pawlik}, A.~H., {Rai{\v c}evi{\'c}}, M., \& {Schaye}, J.
  2013{\natexlab{a}}, \mnras, 430, 2427

\bibitem[{{Rahmati} {et~al.}(2015){Rahmati}, {Schaye}, {Bower}, {Crain},
  {Furlong}, {Schaller}, \& {Theuns}}]{rahm15}
{Rahmati}, A., {Schaye}, J., {Bower}, R.~G., {et~al.} 2015, \mnras, 452, 2034

\bibitem[{{Rahmati} {et~al.}(2013{\natexlab{b}}){Rahmati}, {Schaye}, {Pawlik},
  \& {Rai{\v c}evi{\'c}}}]{rahm13a}
{Rahmati}, A., {Schaye}, J., {Pawlik}, A.~H., \& {Rai{\v c}evi{\'c}}, M.
  2013{\natexlab{b}}, \mnras, 431, 2261

\bibitem[{{Rees} \& {Ostriker}(1977)}]{rees77}
{Rees}, M.~J., \& {Ostriker}, J.~P. 1977, \mnras, 179, 541

\bibitem[{{Richter} {et~al.}(2016){Richter}, {Nuza}, {Fox}, {Wakker}, {Lehner},
  {Ben Bekhti}, {Fechner}, {Wendt}, {Howk}, {Muzahid}, {Ganguly}, \&
  {Charlton}}]{rich16}
{Richter}, P., {Nuza}, S.~E., {Fox}, A.~J., {et~al.} 2016, ArXiv e-prints,
  arXiv:1611.07024

\bibitem[{{Rosas-Guevara} {et~al.}(2015){Rosas-Guevara}, {Bower}, {Schaye},
  {Furlong}, {Frenk}, {Booth}, {Crain}, {Dalla Vecchia}, {Schaller}, \&
  {Theuns}}]{rosa15}
{Rosas-Guevara}, Y.~M., {Bower}, R.~G., {Schaye}, J., {et~al.} 2015, \mnras,
  454, 1038

\bibitem[{{Rubin} {et~al.}(2011){Rubin}, {Prochaska}, {M{\'e}nard}, {Murray},
  {Kasen}, {Koo}, \& {Phillips}}]{rubi11}
{Rubin}, K.~H.~R., {Prochaska}, J.~X., {M{\'e}nard}, B., {et~al.} 2011, \apj,
  728, 55

\bibitem[{{Schaye}(2004)}]{scha04}
{Schaye}, J. 2004, \apj, 609, 667

\bibitem[{{Schaye} \& {Dalla Vecchia}(2008)}]{scha08}
{Schaye}, J., \& {Dalla Vecchia}, C. 2008, \mnras, 383, 1210

\bibitem[{{Schaye} {et~al.}(2010){Schaye}, {Dalla Vecchia}, {Booth}, {Wiersma},
  {Theuns}, {Haas}, {Bertone}, {Duffy}, {McCarthy}, \& {van de Voort}}]{scha10}
{Schaye}, J., {Dalla Vecchia}, C., {Booth}, C.~M., {et~al.} 2010, \mnras, 402,
  1536

\bibitem[{{Schaye} {et~al.}(2015){Schaye}, {Crain}, {Bower}, {Furlong},
  {Schaller}, {Theuns}, {Dalla Vecchia}, {Frenk}, {McCarthy}, {Helly},
  {Jenkins}, {Rosas-Guevara}, {White}, {Baes}, {Booth}, {Camps}, {Navarro},
  {Qu}, {Rahmati}, {Sawala}, {Thomas}, \& {Trayford}}]{scha15}
{Schaye}, J., {Crain}, R.~A., {Bower}, R.~G., {et~al.} 2015, \mnras, 446, 521

\bibitem[{{Shull} {et~al.}(2012){Shull}, {Smith}, \& {Danforth}}]{shul12}
{Shull}, J.~M., {Smith}, B.~D., \& {Danforth}, C.~W. 2012, \apj, 759, 23

\bibitem[{{Springel}(2005)}]{spri05}
{Springel}, V. 2005, \mnras, 364, 1105

\bibitem[{{Steidel} {et~al.}(2011){Steidel}, {Bogosavljevi{\'c}}, {Shapley},
  {Kollmeier}, {Reddy}, {Erb}, \& {Pettini}}]{stei11}
{Steidel}, C.~C., {Bogosavljevi{\'c}}, M., {Shapley}, A.~E., {et~al.} 2011,
  \apj, 736, 160

\bibitem[{{Stocke} {et~al.}(1991){Stocke}, {Case}, {Donahue}, {Shull}, \&
  {Snow}}]{stoc91}
{Stocke}, J.~T., {Case}, J., {Donahue}, M., {Shull}, J.~M., \& {Snow}, T.~P.
  1991, \apj, 374, 72

\bibitem[{{Stocke} {et~al.}(2013){Stocke}, {Keeney}, {Danforth}, {Shull},
  {Froning}, {Green}, {Penton}, \& {Savage}}]{stoc13}
{Stocke}, J.~T., {Keeney}, B.~A., {Danforth}, C.~W., {et~al.} 2013, \apj, 763,
  148

\bibitem[{{Szomoru} {et~al.}(2011){Szomoru}, {Franx}, {Bouwens}, {van Dokkum},
  {Labb{\'e}}, {Illingworth}, \& {Trenti}}]{szom11}
{Szomoru}, D., {Franx}, M., {Bouwens}, R.~J., {et~al.} 2011, \apjl, 735, L22

\bibitem[{{Taylor} {et~al.}(2016){Taylor}, {Davies}, {J{\'a}chym}, {Keenan},
  {Minchin}, {Palou{\v s}}, {Smith}, \& {W{\"u}nsch}}]{tayl16}
{Taylor}, R., {Davies}, J.~I., {J{\'a}chym}, P., {et~al.} 2016, \mnras, 461,
  3001

\bibitem[{{Tumlinson} {et~al.}(2017){Tumlinson}, {Peeples}, \&
  {Werk}}]{tuml2017}
{Tumlinson}, J., {Peeples}, M.~S., \& {Werk}, J.~K. 2017, \araa, 55, 389

\bibitem[{{van de Voort} \& {Schaye}(2013)}]{vand13}
{van de Voort}, F., \& {Schaye}, J. 2013, \mnras, 430, 2688

\bibitem[{{van de Voort} {et~al.}(2011{\natexlab{a}}){van de Voort}, {Schaye},
  {Booth}, \& {Dalla Vecchia}}]{vand11}
{van de Voort}, F., {Schaye}, J., {Booth}, C.~M., \& {Dalla Vecchia}, C.
  2011{\natexlab{a}}, \mnras, 415, 2782

\bibitem[{{van de Voort} {et~al.}(2011{\natexlab{b}}){van de Voort}, {Schaye},
  {Booth}, {Haas}, \& {Dalla Vecchia}}]{vand11a}
{van de Voort}, F., {Schaye}, J., {Booth}, C.~M., {Haas}, M.~R., \& {Dalla
  Vecchia}, C. 2011{\natexlab{b}}, \mnras, 414, 2458

\bibitem[{{Van Sistine} {et~al.}(2016){Van Sistine}, {Salzer}, {Sugden},
  {Giovanelli}, {Haynes}, {Janowiecki}, {Jaskot}, \& {Wilcots}}]{vans16}
{Van Sistine}, A., {Salzer}, J.~J., {Sugden}, A., {et~al.} 2016, \apj, 824, 25

\bibitem[{{Vargas} {et~al.}(2017){Vargas}, {Heald}, {Walterbos}, {Fraternali},
  {Patterson}, {Rand}, {J{\'o}zsa}, {Gentile}, \& {Serra}}]{varg17}
{Vargas}, C.~J., {Heald}, G., {Walterbos}, R.~A.~M., {et~al.} 2017, \apj, 839,
  118

\bibitem[{{Werk} {et~al.}(2013){Werk}, {Prochaska}, {Thom}, {Tumlinson},
  {Tripp}, {O'Meara}, \& {Peeples}}]{werk13}
{Werk}, J.~K., {Prochaska}, J.~X., {Thom}, C., {et~al.} 2013, \apjs, 204, 17

\bibitem[{{Werk} {et~al.}(2014){Werk}, {Prochaska}, {Tumlinson}, {Peeples},
  {Tripp}, {Fox}, {Lehner}, {Thom}, {O'Meara}, {Ford}, {Bordoloi}, {Katz},
  {Tejos}, {Oppenheimer}, {Dav{\'e}}, \& {Weinberg}}]{werk14}
{Werk}, J.~K., {Prochaska}, J.~X., {Tumlinson}, J., {et~al.} 2014, \apj, 792, 8

\bibitem[{{White} \& {Frenk}(1991)}]{whit91}
{White}, S.~D.~M., \& {Frenk}, C.~S. 1991, \apj, 379, 52

\bibitem[{{Wiersma} {et~al.}(2009){Wiersma}, {Schaye}, \& {Smith}}]{wier09}
{Wiersma}, R.~P.~C., {Schaye}, J., \& {Smith}, B.~D. 2009, \mnras, 393, 99

\bibitem[{{Wisotzki} {et~al.}(2016){Wisotzki}, {Bacon}, {Blaizot},
  {Brinchmann}, {Herenz}, {Schaye}, {Bouch{\'e}}, {Cantalupo}, {Contini},
  {Carollo}, {Caruana}, {Courbot}, {Emsellem}, {Kamann}, {Kerutt}, {Leclercq},
  {Lilly}, {Patr{\'{\i}}cio}, {Sandin}, {Steinmetz}, {Straka}, {Urrutia},
  {Verhamme}, {Weilbacher}, \& {Wendt}}]{wiso16}
{Wisotzki}, L., {Bacon}, R., {Blaizot}, J., {et~al.} 2016, \aap, 587, A98

\bibitem[{{Wisotzki} {et~al.}(2018){Wisotzki}, {Bacon}, {Brinchmann},
  {Cantalupo}, {Richter}, {Schaye}, {Schmidt}, {Urrutia}, {Weilbacher},
  {Akhlaghi}, {Bouch{\'e}}, {Contini}, {Guiderdoni}, {Herenz}, {Inami},
  {Kerutt}, {Leclercq}, {Marino}, {Maseda}, {Monreal-Ibero}, {Nanayakkara},
  {Richard}, {Saust}, {Steinmetz}, \& {Wendt}}]{wiso18}
{Wisotzki}, L., {Bacon}, R., {Brinchmann}, J., {et~al.} 2018, \nat, 562, 229

\bibitem[{{Zhang} \& {Zaritsky}(2017)}]{zhan17}
{Zhang}, H., \& {Zaritsky}, D. 2017, Nature Astronomy, 1, 0103

\bibitem[{{Zhang} {et~al.}(2018){Zhang}, {Abraham}, {van Dokkum}, {Merritt}, \&
  {Janssens}}]{zhan18}
{Zhang}, J., {Abraham}, R., {van Dokkum}, P., {Merritt}, A., \& {Janssens}, S.
  2018, \apj, 855, 78

\end{thebibliography}

\appendix
\subsection{A1. CONVERGENCE TEST}
In this section, we present a test to verify the convergence of the predictions from the simulation with different numerical resolutions.  We compare emission maps from four simulations:  \textsc{ref376}, \textsc{ref752}, \textsc{recal752}, and \textsc{ref1504}.  \textsc{ref376} and \textsc{ref1504} are `intermediate resolution' simulations, with initial baryonic particle masses of 1.81$\times10^6$~M$_{\odot}$ and gravitational softening lengths of 0.70 proper kpc while \textsc{ref752} and \textsc{recal752} are `high resolution' simulations, with initial baryonic particle masses of  2.26$\times10^5$~M$_{\odot}$ and gravitational softening lengths of 0.35 proper kpc.  \textsc{ref1504} is the simulation used in the main analysis of this paper and has a box size of 100 comoving Mpc, whereas the other three simulations have box sizes of 25 comoving Mpc. 

\begin{figure*}
\centering
\includegraphics[width=0.8\linewidth]{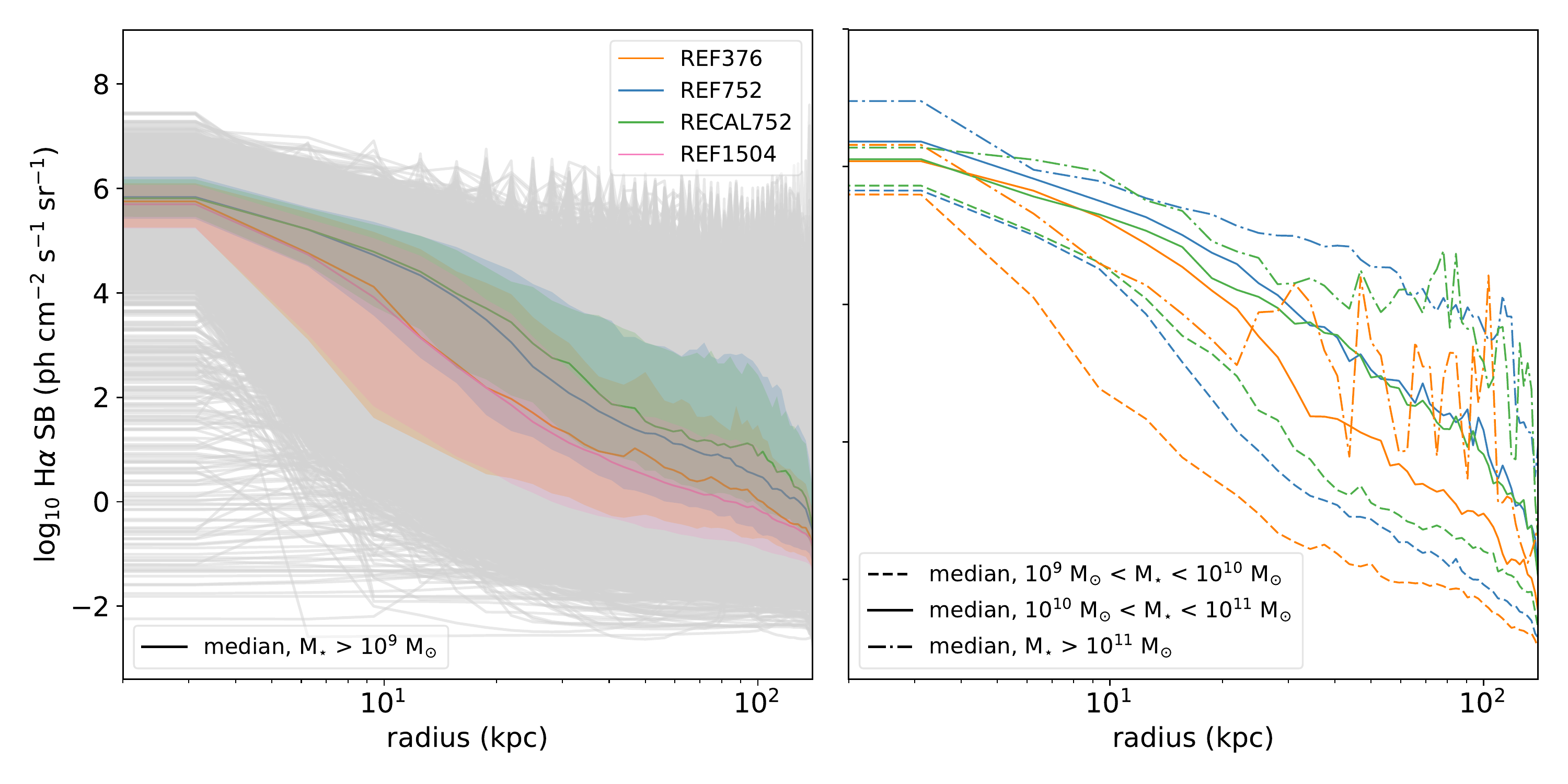}
\caption{Comparison of H$\alpha$ azimuthally averaged surface brightness profiles from four EAGLE runs: \textsc{ref376}, \textsc{ref752}, \textsc{recal752}, and \textsc{ref1504}. \textsc{ref752} and \textsc{recal752} are higher resolution runs with box size of 25 cMpc.  \textsc{ref376} and \textsc{ref1504} are lower resolution simulations with box sizes of 25 cMpc and 100 cMpc, respectively (\textsc{ref1504} is the simulation used in the main analysis of this paper).  In the left panel, the median H$\alpha$ surface brightness profiles for galaxies with stellar mass M$_{\star} > 10^9$~M$_{\odot}$ are shown for each simulation, with same-colored shading filling the area between the 25th and 75th percentiles.
In the right panel, the H$\alpha$ surface brightness profiles for galaxies in three mass bins (10$^9$ M$_{\odot}$ -- 10$^{10}$ M$_{\odot}$, 10$^{10}$ M$_{\odot}$ -- 10$^{11}$ M$_{\odot}$, and 10$^{11}$ M$_{\odot}$ and up) are shown for each simulation.  For clarity, the \textsc{ref1504} simulation is left out of the right panel plot.
The profiles are consistent with one another within the scatter, though the higher resolution runs trend to higher surface brightnesses.}
\label{Fig25Mpc}
\end{figure*}

Fig.~\ref{Fig25Mpc} shows the H$\alpha$ azimuthally averaged surface brightness profiles for each simulation.  
Although the differences between the median profiles of the different simulations are consistent with each other within the galaxy-to-galaxy scatter, there are significant variations particularly at large radii, where the higher resolution simulations predict higher surface brightnesses for each mass bin.
We hypothesize that the higher resolution runs introduce more density peaks into the simulation, which boosts the H$\alpha$ emission; hence the results based on the 100 cMpc box used throughout are conservative with regards to the detectability of the CGM.

\subsection{A2. CONTRIBUTION OF SOURCES OF EMISSION}

In this section, we calculate the emission maps from the EAGLE simulation from separate sources of emission and compare the results. In the first case, we calculate the emission from all gas particles including star-forming particles, using the prescription outlined in Section 2, where the non-star-forming gas particles are assumed to be optically thin to the metagalactic ionizing radiation (this is the prescription used in the analysis of the main part of this paper).  In the second case, we include emission from all non-star-forming gas particles (leaving out star-forming gas).  In the third case, we apply the prescription outlined in \citet{rahm13} to estimate the effects of self-shielding in the simulation.  We use Equations A1 and A8 from the analysis of \citet{rahm13} to calculate the neutral fraction in each non-star-forming gas particle and omit this fraction from the emission calculation.  This is a conservative estimate of the effects of self-shielding, since we completely remove the optically thick fraction of gas from the particle (neglecting that a fraction of this gas in the outer shell will also see ionizing radiation and emit H$\alpha$ emission).  To correctly account for the fluorescent H$\alpha$ emission of optically thick clouds it is necessary to perform a full radiative transfer calculation, which is beyond the scope of this work.
These cases thus reflect the range in which we expect the H$\alpha$ emission to lie in reality.  
It is important to note that for the second and third prescriptions applied here, we ignore the effect of ionizing radiation from local sources, such as young stars and quasars, which are expected to become important for the gas particles at densities where self-shielding from the UVB comes into play as demonstrated by the first case, i.e.\ emission from local sources may boost the emission even more where the emission drops due to self-shielding \citep{rahm13a}.

\begin{figure*}
\centering
\includegraphics[width=1.0\linewidth]{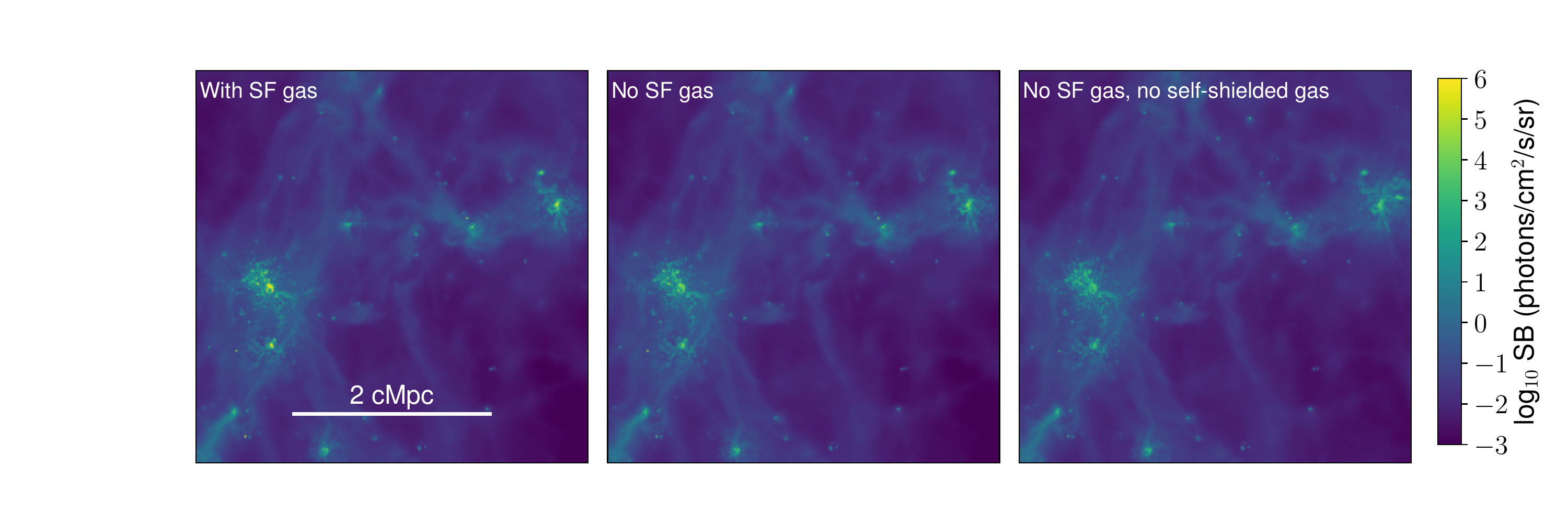}
\caption{Comparison of H$\alpha$ emission maps projected from the EAGLE simulation when including different sources of emission.  The left panel shows the emission map including emission from all gas particles (star-forming and non-star-forming) and assuming all non-star-forming gas particles are optically thin to the metagalactic ionizing radiation.  The middle panel shows the emission map including emission from only non-star-forming gas particles and assuming these particles are optically thin to the metagalactic ionizing radiation.  The right panel shows the emission map excluding emission from optically thick gas \citep[i.e. excluding star-forming gas and self-shielded gas, where the self-shielded fraction was estimated following][]{rahm13}.  }
\label{FigEmissionSourceMaps}
\end{figure*}

\begin{figure}
\centering
\includegraphics[width=0.4\linewidth]{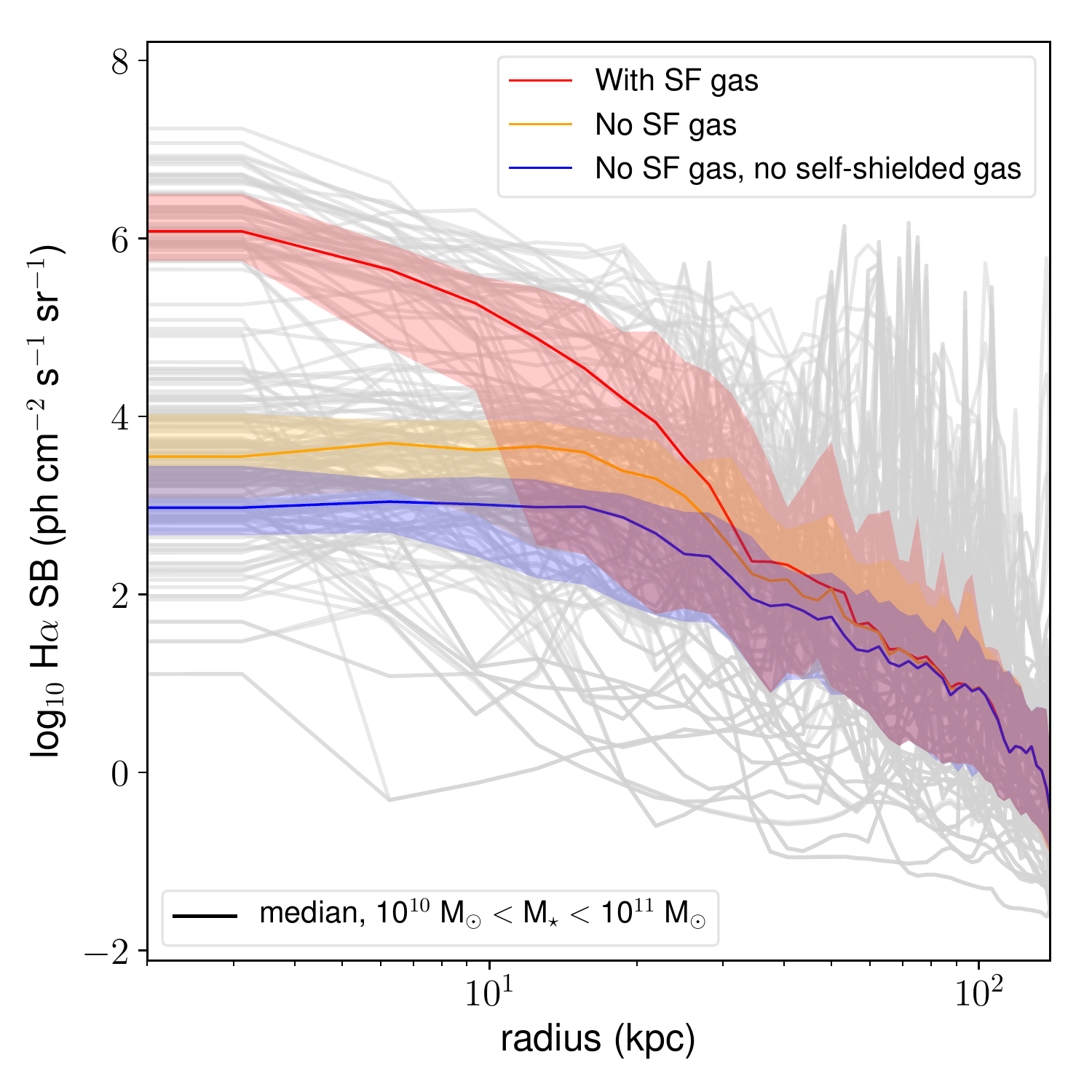}
\caption{Comparison of H$\alpha$ azimuthally averaged surface brightness profiles from the EAGLE simulation when including different sources of emission (same as Fig.~\ref{FigEmissionSourceMaps}).
Although the effect of local sources of ionizing radiation, and to a lesser extent self-shielding, is large within galaxies, the predictions of the different prescriptions converge in the circumgalactic medium. }
\label{FigEmissionSourceProfiles}
\end{figure}

In Fig.~\ref{FigEmissionSourceMaps}, we show the emission maps for each case side-by-side.  By eye, it is difficult to see any difference in these maps, except in the central regions of galaxies.
In Fig.~\ref{FigEmissionSourceProfiles}, we compare the H$\alpha$ azimuthally averaged surface brightness profiles for each case.  The first case differs from the second and third cases by more than two orders of magnitude at radii inside the galaxy, whereas at the inner edge of the CGM, they converge to similar values (i.e.\ star-forming and self-shielded gas is concentrated within the galaxies).  
For comparison, the H$\alpha$ surface brightness for optically thick clouds arising from recombination following photoionization by the UV background can be computed following Equation 5 of \citet{goul96}, where $\phi(\nu)$ is given by the \citet{haar01} spectrum and $\eta_{\mathrm{thick}}$ is the number of H$\alpha$ photons per incident ionizing photon for Case B recombination (assuming a gas temperature of 10$^4$ K yields $\eta_{\mathrm{thick}}$ = 0.45)\footnote{Note that this calculation differs from the simulation in that it assumes Case B rather than Case A recombination, a constant gas temperature of 10$^4$ K, and ignores emission processes other than recombination.  The calculation is consistent with the simulation in that dust is ignored.}.  This results in an H$\alpha$ surface brightness of $\approx1\times10^3$ photons s$^{-1}$ cm$^{-2}$ sr$^{-1}$.  This estimate agrees well with the maximum surface brightness of the third case considered here (the blue line in Fig~\ref{FigEmissionSourceProfiles}); while the third case includes only optically thin gas, at its maximum surface brightness the emission is coming from clouds that are very close to being optically thick, so this agreement is reassuring.
We conclude that although the effect of local sources of ionizing radiation, and to a lesser extent self-shielding, is large within galaxies, the predictions of the different prescriptions converge in the CGM.

\begin{figure*}
\centering
\includegraphics[width=1.0\linewidth]{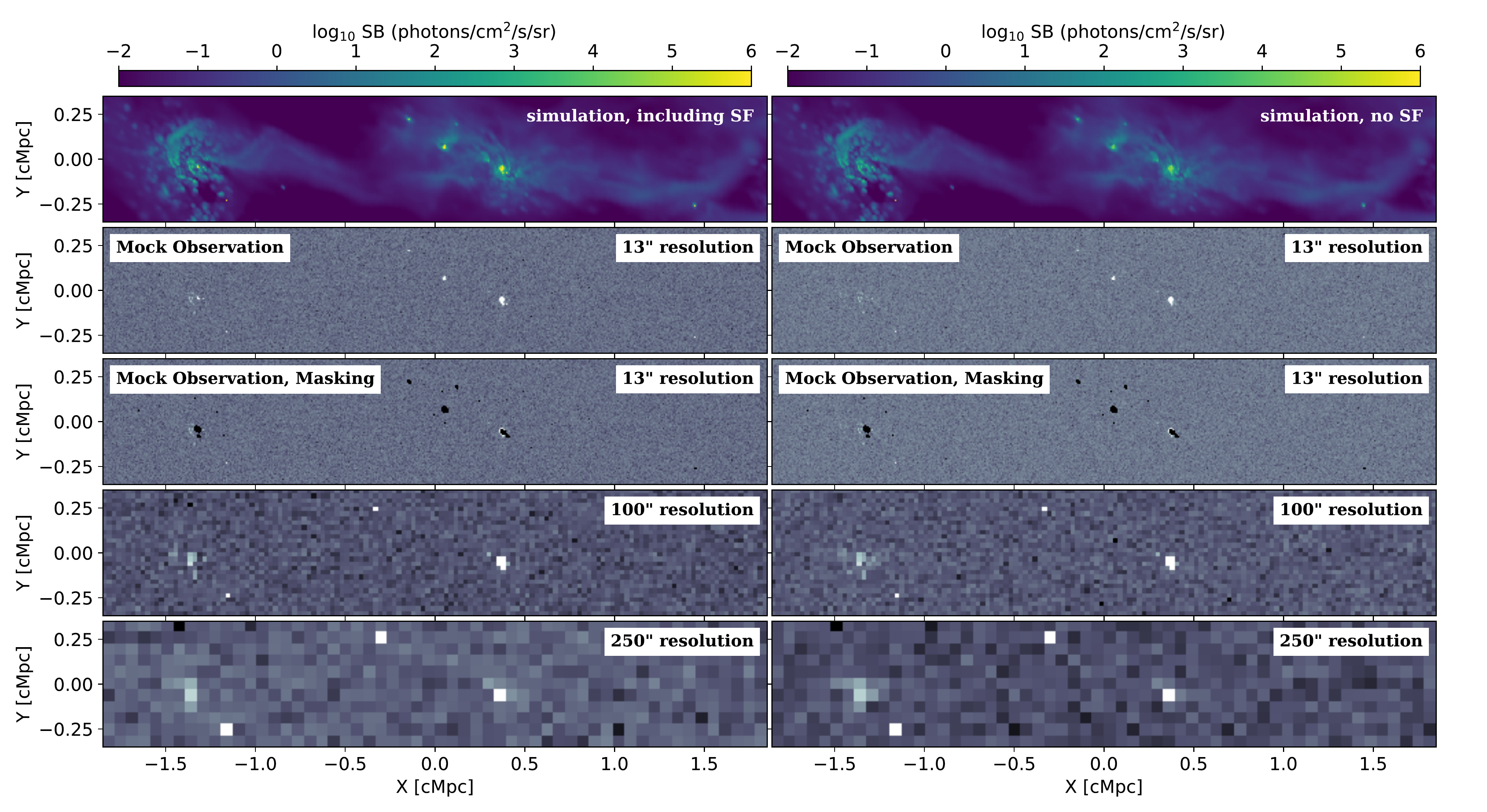}
\caption{(Left column of plots redisplayed from Fig.~\ref{FigBinning}.) Same as Fig.~\ref{FigBinning}.  Comparison of the emission from the CGM and IGM including the star-forming gas (in the left column) to emission omitting star-forming gas (right column).  It is clear that even when omitting star-forming gas to isolate extragalactic gas emission from galactic emission there are still significant sources of emission.}
\label{FigIGMSFComparison}
\end{figure*}

In Fig.~\ref{FigIGMSFComparison}, we test the effects of not including star-forming gas particles in the mock observations of the IGM, to isolate emission from the CGM and IGM gas.  We reproduce Fig.~\ref{FigBinning} in the left column of plots and compare this to an identical set of plots in the right column, except that these plots were made without including emission from star-forming particles i.e.\ not including galactic gas.  We compare the mock observations side-by-side, with star-forming gas included (omitted) in the left (right) column of plots.  One can see that even when sources of emission within the galaxies are omitted (in the right hand panels), the emission left over from the CGM and IGM is still significant.

\clearpage
\end{document}